\documentclass[aps,prl,preprint,groupedaddress]{revtex4}
\usepackage{graphicx} 
\def \pt {{p_\bot}}
\def \mt {{m_\bot}}
\def \pl {{p_\parallel}}
\def \e {{\rm e}}
\def \eq {{\rm e}_q}
\def \d {{\rm d}}
\def \Q {{\bf Q}}
\def \q {{\bf q}}
\def \Z {{\cal Z}}
\def \Zq {{\cal Z}_q}
\begin{document}
\title{Large Transverse Momenta in Statistical Models of
High Energy Interactions.}

\author{T. Wibig}
\affiliation{Experimental Physics Dept., University of \L \'{o}d\'{z}
Pomorska 149/153, 90--236 \L \'{o}d\'{z}}
\affiliation{ The Andrzej So\l tan Institute For Nuclear Studies,
Cosmic Ray Lab., \L \'{o}d\'{z}, Uniwersytecka 5, POB 447, \L \'{o}d\'{z} 1 ;Poland}
\author{I. Kurp}
\affiliation{The Andrzej So\l tan Institute For Nuclear Studies,
Cosmic Ray Lab., \L \'{o}d\'{z}, Uniwersytecka 5, POB 447, \L \'{o}d\'{z} 1 , Poland}

\date{\today}

\begin{abstract}
The creation of particles with large transverse momenta in high energy hadronic collisions
is a long standing problem. The transition from small- (soft) to hard- parton
scattering `high-$\pt$' events is rather smooth. In this paper we apply
the non-extensive statistical framework to calculate transverse momentum distributions 
of long lived hadrons created at energies from low
($\sqrt{s}\approx 10$ GeV) to the highest energies 
available in collider experiments ($\sqrt{s}\approx 2000$ GeV). Satisfactory agreement
with the experimental data is achieved.
The systematic increase of the non-extensivity parameter with energy found
can be understood as phenomenological evidence
for the increased role of long range correlations in the hadronization process.

Predictions concerning the rise of average transverse momenta up to the highest cosmic ray energies
are also given and discussed.

\end{abstract}
\maketitle

\section{Introduction}
In what follows we write the momentum, $p$, in units of energy (e.g. GeV);
strictly, of course it should be in units of energy divided by $c$.

For soft (low $\pt$ events) particle production, the QCD-based string model 
predicts that the transverse momentum (transverse mass: $\mt=\sqrt{\pt^2+m^2}$)
distribution of the produced quarks should be, in general, of the form 
\begin{equation}
{\d \sigma \over \d \pt^2 }\sim \e ^{-\pi \mt^2/\kappa^2}~,
\end{equation}
where $\kappa$ is the string tension. According to \cite{Bialas:1999zg},
when the string tension may fluctuate, the string hadronization becomes consistent with 
thermal behaviour, and
\begin{equation}
{\d \sigma \over \d \pt^2 }\sim \e ^{-\pi \mt/T}.
\end{equation}
This means that in the soft (small $\pt$) region, the partonic string 
fragmentation picture, which is based on first principles, can be, to some extent,
successfully replaced by the ``phenomenological'' statistical model. 

Much more interesting, and complex, is the case of high transverse momentum physics.

The main reason for the failure of the traditional thermodynamical models of multi-particle production 
\cite{Fermi:1950jd}, \cite{Hagedorn:1970gh},\cite{Hagedorn:1972sk} was 
the experimentally observed significant increase of the production of particles with 
high transverse momenta discovered 
in the mid-seventies in the Intersecting Storage Ring (ISR) experiments. 
Its interpretation in the framework of jet models \cite{Feynman:1977yr,Field:1978fa} 
strongly supported the parton idea.
But then new problems arose. 
The ``hard'' parton scattering expected from field theories implies
roughly that the momenta should fall off as $\pt\!^{-4}$ 
\begin{equation}
{\d \sigma \over \d \pt^2}~\sim~F_A(x_a,\:q_\bot^2)\:F_B(x_b,\:q_\bot^2)\:
{\alpha_s^2(q_\bot^2) \over q_\bot^4}
\label{hardpt}
\end{equation}
(where the $F$'s are the respective structure functions).
However the power-law fit to the data 
is closer to $\pt\!^{-8}$, a fact that was realized already in, 
e.g.,  \cite{Field:1977ve} and \cite{Feynman:1978dt}.

The commonly accepted explanation of high $\pt$ behaviour in inclusive $pp$ (and $p\bar p$)
spectra
is that at high enough energies the QCD hard scattering
effects start to play a significant role. The transition 
''soft''$\rightarrow$''not-so-hard''$\rightarrow$''hard'' {\em is believed} to combine 
the exponential
low-$\pt$ domain with the asymptotic $\pt\!^{-4}$ tail in such a way that the
expected distributions are, around a few GeV/c and for the interaction energies available at present, 
just as observed.
To follow this idea in detail further 
QCD calculations are needed,
and to cover the whole $\pt$ range some 
interplay of perturbative (high $\pt$ - ``hard'') and
non-perturbative (``soft'') physics have to be developed in a self-consistent 
way. The problem is not trivial and here we point a way forward.

In the present paper we will show that
the problem can be quite satisfactorily solved by way of  
statistical language as an effect of multi-particle, long-range correlations. 
We {\em believe} that both descriptions are 
equivalent. 

In the standard statistical (thermodynamical) models of hadronization, high
$\pt$'s can be explained (?) only by introducing additional mechanisms.
Two such attempts have appeared recently. 

One way (\cite{Becattini:2000pr}) is to smear the $\pt$s with an additional, dynamic, term 
describing the collective motion
of different parts of  a ``pre-hadronizing'' state of matter. It can be achieved by introducing
a distribution 
in momentum space of
many fireballs created in the collision. However this mechanism 
cannot be responsible for very high $\pt$
tails. 

\begin{figure}[ht]
\centerline{
\includegraphics[width=6cm]{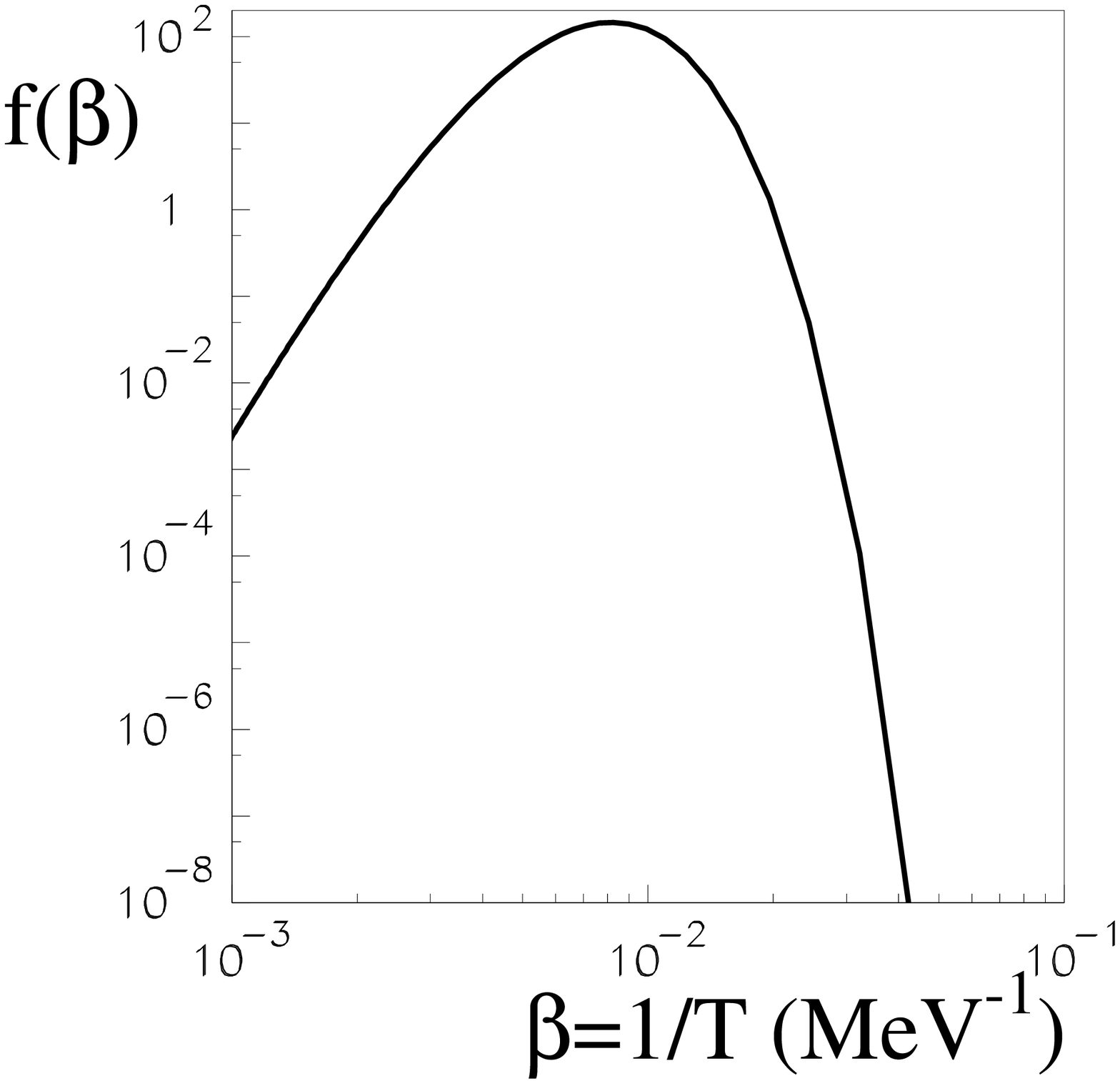}
\includegraphics[width=6cm]{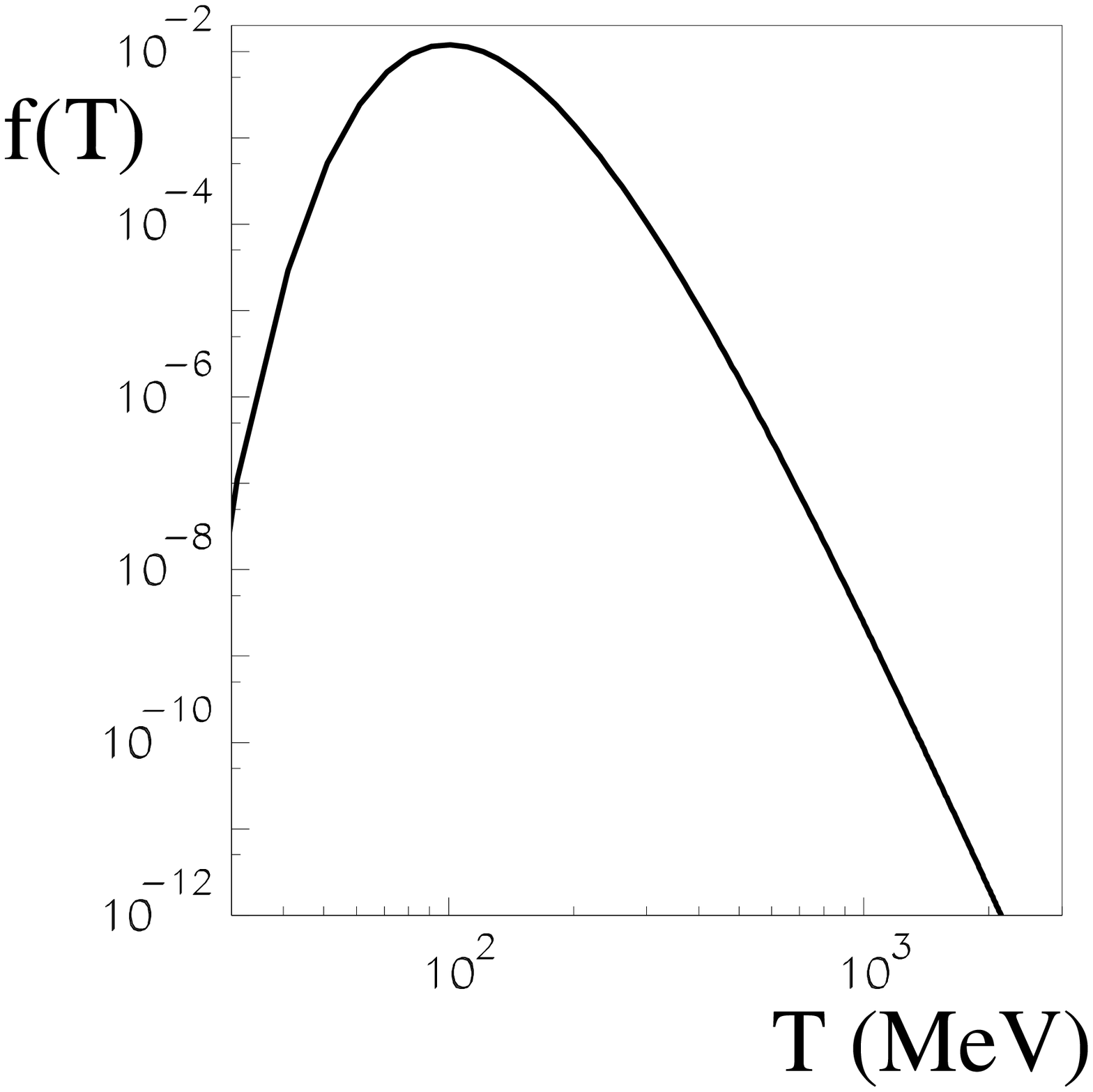}}
\caption{Distribution of $\beta=1/T$ (left) and $T$ (right) in the
picture of $\pt$ - broadening by temperature fluctuations with $\lambda=10$ and $T_0=
\left\langle 1/T \right \rangle^{-1}=\alpha/\lambda=110$ MeV.
\label{jedennadt}}
\end{figure}

The second idea (\cite{Beck:2001ts,Wilk:2001db,Wilk:2002ic})
  is to allow the Hagedorn temperature to
fluctuate around its mean. A very good description of the data can be 
obtained assuming that \cite{Wilk:2001db,Wilk:2002ic}
\begin{equation}
f(\beta)=p(1/T)={\alpha^\lambda \over \Gamma(\lambda)}
\left({1 \over T } \right)^{\lambda-1}
\exp \left( - {\alpha \over T}\right)
\end{equation}
with parameters
\begin{equation}
\left \langle {1 \over T}\right \rangle ={\lambda \over \alpha}~~,~~~~~
\left \langle {1 \over T^2}\right \rangle -
\left \langle {1 \over T}\right \rangle ^2
={\lambda \over \alpha^2}~.
\end{equation}
It is clear that the relatively narrow $\Gamma$ distribution in $\beta=1/T$ gives necessary 
very substantial
tails in the distribution of $T$. The situation is presented in Fig.~\ref{jedennadt}. In
fact, interesting cases with high $\pt$ originate in events with
actual Hagedorn temperatures much greater than the ``hadronic 
soup boiling temperature'' \cite{soup}. This is
rather inconsistent with the general Hagedorn fireball picture.

In the present work we modify the classical Hagedorn idea by introducing the
long-range correlations in statistical way.

\section{A Standard statistical model}
One way to describe the statistical properties of the system 
is by introducing the concept of the partition function.
Its classical definition 
for the system in the state described by the vector $\Q_0$ 
({\it canonical} statistical ensemble)
is given by
\begin{equation}
\Z(\Q)~=~\sum \limits_{\rm states} \e^{-E/T} \:\delta _{\Q,\Q_0} ,
\label{zdef}
\end{equation}
where $E$ is the energy of the system. For the purposes of the present paper,
where the system contains hadrons created in high energy hadronic collisions,
$Q$ can be  limited to a three dimensional
vector $\Q$ and its components are the total
charge, the baryon number and the strangeness of the system $(Q,\:B,\:S)$.
Of course, when needed, a generalization is straightforward.

We follow here the formalism developed by Becattini (see, e.g., \cite{Becattini:1997rv}):

By using the integral representation
of the Kronecker $\delta$ factor, Eq.(\ref{zdef}) becomes
\begin{equation}
\Z(\Q)~=~\sum \limits_{\rm states}
\:{1 \over \left(2\pi \right)^3}
\int \limits_{0}^{2\pi}
\int \limits_{0}^{2\pi}
\int \limits_{0}^{2\pi}
\d^3 \phi \:
\e^{-E/T} \:
\e^{i (\Q_0-\Q)\:\phi} ~.
\label{zdefc}
\end{equation}
If we have a quantum gas with $N_b$ boson and $N_f$ fermion species 
a summation over states can be performed, giving
\begin{equation}
\Z(\Q)= {1 \over \left(2\pi \right)^3}
\int
\d^3 \phi  
\e^{i \Q_0\:\phi} 
\exp\left[
\sum \limits_{N_b}
\sum \limits_{k}
\log \left( 1-\e^{-{E_k\over T} - i \q_k\phi }\right)^{-1}+
\sum \limits_{N_f}
\sum \limits_{k}
\log \left( 1+\e^{-{E_k \over T} - i \q_k\phi }\right)\right].
\label{zsumk}
\end{equation}
The sum over phase space 
cells, $k$, can be, in the continuous limit, replaced by an integration over
momentum space:
\begin{equation}
\sum \limits_{k} ~\longrightarrow ~(2J_i+1) 
{ V \over \left(2\pi \right)^3 }\:
\int \d ^3 p .
\label{sum2int}
\end{equation}

The average multiplicity of the $i^{\rm th}$ hadron type 
can be obtained from the partition function by
introducing the fictitious fugacity factor $\lambda_i$
\begin{equation}
\langle n_i \rangle~=~ \left.{\partial \over \partial \lambda_i}  \log
\left({\Z(\Q_0,\:\lambda_i) }\right) \right|
_{\lambda_i=1} ~,
\label{multdef}
\end{equation}
thus
\begin{equation}
\langle n_i \rangle~=~
(2J_i+1) { V \over \left(2\pi \right)^3 }\:
\:{1 \over \left(2\pi \right)^3}
\int \limits_{0}^{2\pi}
\int \limits_{0}^{2\pi}
\int \limits_{0}^{2\pi}
\d^3 \phi \:
\int \d ^3 p
\:\left[ \e^{E/T}\:\e^{i \:\q_i \phi} \pm 1 \right]^{-1}~,
\label{mult1}
\end{equation}
where the upper sign is for fermions and the lower is for bosons.
Because the $\e^{-E/T}$ factor is expected to be small for 
all particles except pions
($T\approx 100$ MeV) the following approximation can be made 
in such cases 
\begin{equation}
{1 \over \e^{E/T}\:\e^{i \:\q_i \phi} \pm 1 }~ \longrightarrow ~
\e^{-E/T - i \:\q_i \phi}
\label{approx1}
\end{equation}
and then
\begin{eqnarray}
\langle n_i \rangle~\approx~ {1 \over \Z(\Q_0)}
\:{1 \over \left(2\pi \right)^3}
\int \limits_{0}^{2\pi}
\int \limits_{0}^{2\pi}
\int \limits_{0}^{2\pi}
\d^3 \phi \:
\Z(\Q_0)\:\e^{-i \:\q_i \phi}
~(2J_i+1) { V \over \left(2\pi \right)^3 }\:
\int \d ^3 p
\:\:\e^{-E/T}~=~\nonumber
\\~~~~~~~~~~~~~=~
{\Z(\Q_0 -\q_i) \over \Z(\Q_0)}
\:(2J_i+1) { V \over \left(2\pi \right)^3 }\:
\int \d ^3 p
\:\e^{-E/T}~.
\label{mult2}
\end{eqnarray}
Very good agreement with the measured particle ratios was found in, e.g., 
\cite{Becattini:1997rv}.
Eqs.(\ref{mult1}, \ref{mult2}) can also be used to calculate the respective 
transverse momentum distributions for
pions and heavier hadrons produced in the thermodynamical hadronization
process.

The high $\pt$ tails of these distributions are known \cite{Hagedorn:1970gh} to fall like
\begin{equation}
{f(\pt)~~} _{\overrightarrow {\pt \gg m, T}}~~
\pt^{3/2}\:\e^{-\pt/T}~,
\label{asymp}
\end{equation}
what is in agreement with the low energy experimental results but there is a
clear underestimate of the
high $\pt$'s at collider energies.

\section{Modifications of the statistical hadronization model}
The statistical way is ``phenomenological'' in a sense. 
The conventional Boltzmann-Gibbs
description shown above should be modified, and a new parameter 
describing the correlation
``strength'', however defined, ought to be introduced. 
Of course, in the limit of the absence of correlations the 
new description should approach the Boltzmann form.

From the theoretical point of view there could be
infinitely many ``generalized'' statistics.
Following `Ockham razor' we should choose
that which is simple and has a clear 
theoretical background. In the present paper we
test the possibility, proposed by Tsallis \cite{Tsallis:1988eu},
based on the modification of the 
classical entropy definition
\begin{equation}
S_{\rm BG}~=~-k \sum \limits_{i}^{W} \:p_i\ln p_i
\label{entrobg}
\end{equation}
of the form
\begin{equation}
S_{q}~=~k {1-\sum \limits_{i}^{W} \:p_i^q \over q-1}~
\label{entrots}
\end{equation}
introducing the new parameter $q$, the non-extensivity parameter.
This modification has been adopted in
other physical applications (see, e.g., \cite{Beck:2000nz}).

Maximization of the entropy
requirement
with the 
total energy constraint
\begin{equation}
{\sum \limits_{i} \:p_i^q E_i\over
\sum \limits_{i} \:p_i^q}~=~E_0
\end{equation}
leads to the probability given by
\begin{equation}
p_i^q~=~{1\over Z_q}\:
\left[1-(1-q)/T_q(E_i-E_0) \right]^{q/(1-q)}~,
\label{peq0}
\end{equation}
where $Z_q$ is the normalization constant related to $\Z(q)$
of Eq.(\ref{zdef}) and the Boltzmann
terms are replaced by the probabilities of the form given in Eq.(\ref{peq0}).

It can be mentioned here, that the name of this generalization:
``the non-extensive statistics'' comes from the fact that the 
entropy $S_{q}$ defined by Eq.(\ref{entrots}), in opposition to the 
$S_{\rm BG}$, is
non-extensive parameter (the entropy of the system consisting of two 
separated parts is not a sum of their entropies). A similar statement
is valid for
the total energy of the system. If it consists of $n$ isolated
parts (e.g., gas of ``non-interacting'' hadrons), each of energy $E_i$,
the system total energy
is not equal to $\sum_i^n E_i$ but given by
\begin{equation}
E~=~
\sum \limits_i^n E_i~+~
(q-1)/T
\sum \limits_{i,j} E_i E_j~+~
(q-1)^2/T^2
\sum \limits_{i,j,k} E_i E_j E_k~+~\dots~.
\end{equation}
Additional terms can be interpreted as an effect of the
long-range correlation which appears to be not only the
mathematical construction adopted here, but has something to do with the
real cause of the correlation.

Equation (\ref{peq0}) can be rewritten introducing new
symbol, $\eq$, defined as
\begin{equation}
(\eq)^x  =
\left[ 1+(1-q)x\right]^{q/(1-q)}
\end{equation}
(for completeness, in the q=1 limit we have, as we should, $\e_1^x=\e^x$). Then
\begin{equation}
p_i^q~\sim~
\left[1-(1-q)/T_q(E_i-E_0) \right]^{q/(1-q)}
\sim
\eq ^{-E_i/T_q}
\label{peq}
\end{equation}
and the partition function can be written in the form
\begin{equation}
\Zq(\Q)~=~\sum \limits_{\rm states}
\:{1 \over \left(2\pi \right)^3}
\int \limits_{0}^{2\pi}
\int \limits_{0}^{2\pi}
\int \limits_{0}^{2\pi}
\d^3 \phi \:
\eq^{-E/T} \:
\e^{i (\Q_0-\Q)\:\phi} ~.
\label{zqdef}
\end{equation}

With such a modification, the transverse momentum distribution becomes
\begin{eqnarray}
f(\pt)~\sim~
{\Zq(\Q_0 -\q_i) \over \Zq(\Q_0)}
\:(2J_i+1) { V \over \left(2\pi \right)^3 }\:
\int \d \pl \:\pt
\:\eq^{-E/T}
\label{multq}
\end{eqnarray}
for particles other than pions, while for pions the modified Eq.(\ref{mult1}) should be used:
\begin{equation}
f(\pt)~\sim~
(2J_i+1) { V \over \left(2\pi \right)^3 }\:
\:{1 \over \left(2\pi \right)^3}
\int \limits_{0}^{2\pi}
\int \limits_{0}^{2\pi}
\int \limits_{0}^{2\pi}
\d^3 \phi \:
\int \d \pl \:\pt
\:\left[ \eq^{E/T}\:\e^{i \:\q_i \phi} - 1 \right]^{-1}~,
\label{multpiq}
\end{equation}
(the vector $\q_i$ in the above equations represents
the $i^{\rm th}$ particle type and $q$ is the non-extensivity parameter).

\begin{figure}
\centerline{
\includegraphics[width=8.5cm]{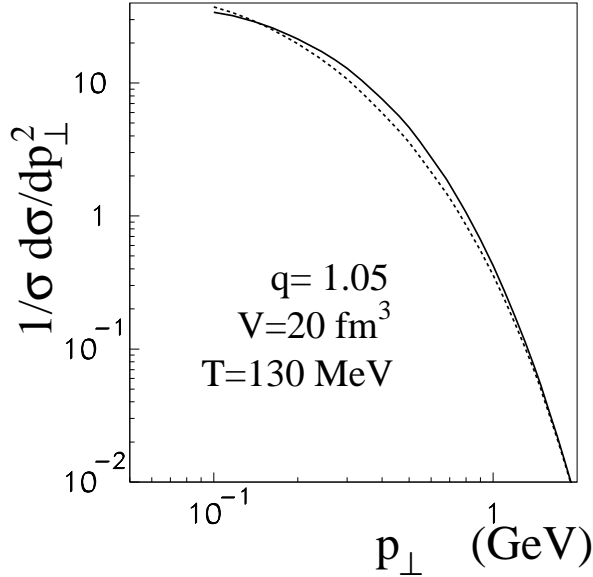}}
\caption{Distributions of $\pt$ for charged primary produced hadrons
(dashed line) and for long-lived hadrons after all decays (solid line)
($T$=130 MeV, $V$=20 fm$^3$ and $q$=1.05). (Strictly, the abscissa, $\pt$, should be in units of GeV/$c$
but we disregard the '$c$' here and elsewhere.)
\label{nodec}}
\end{figure}

In the present work we have evaluated $\Zq$ functions for a variety of its parameters
$T$, $V$, and $q$, and for $\Q$ values which cover the production of over 100 hadrons 
of masses below 2 GeV. The decays of short-lived particles were then performed (with three-body
decay products distributed uniformly in the Dalitz plot). The effect on the $\pt$ distribution caused by
the decays is shown in 
Fig.~\ref{nodec}

\begin{figure}
\centerline{
\includegraphics[width=5.5cm]{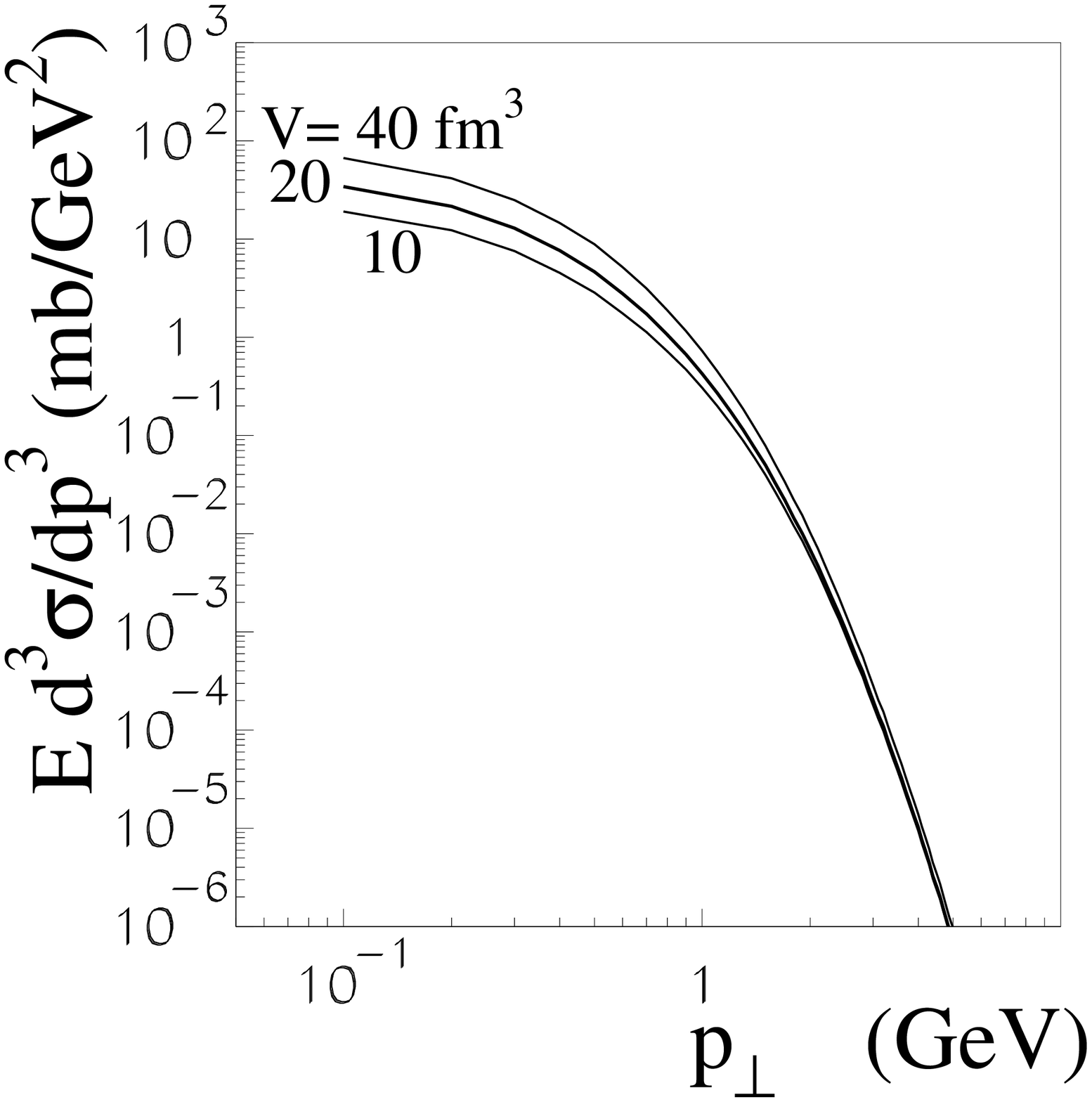}
\includegraphics[width=5.5cm]{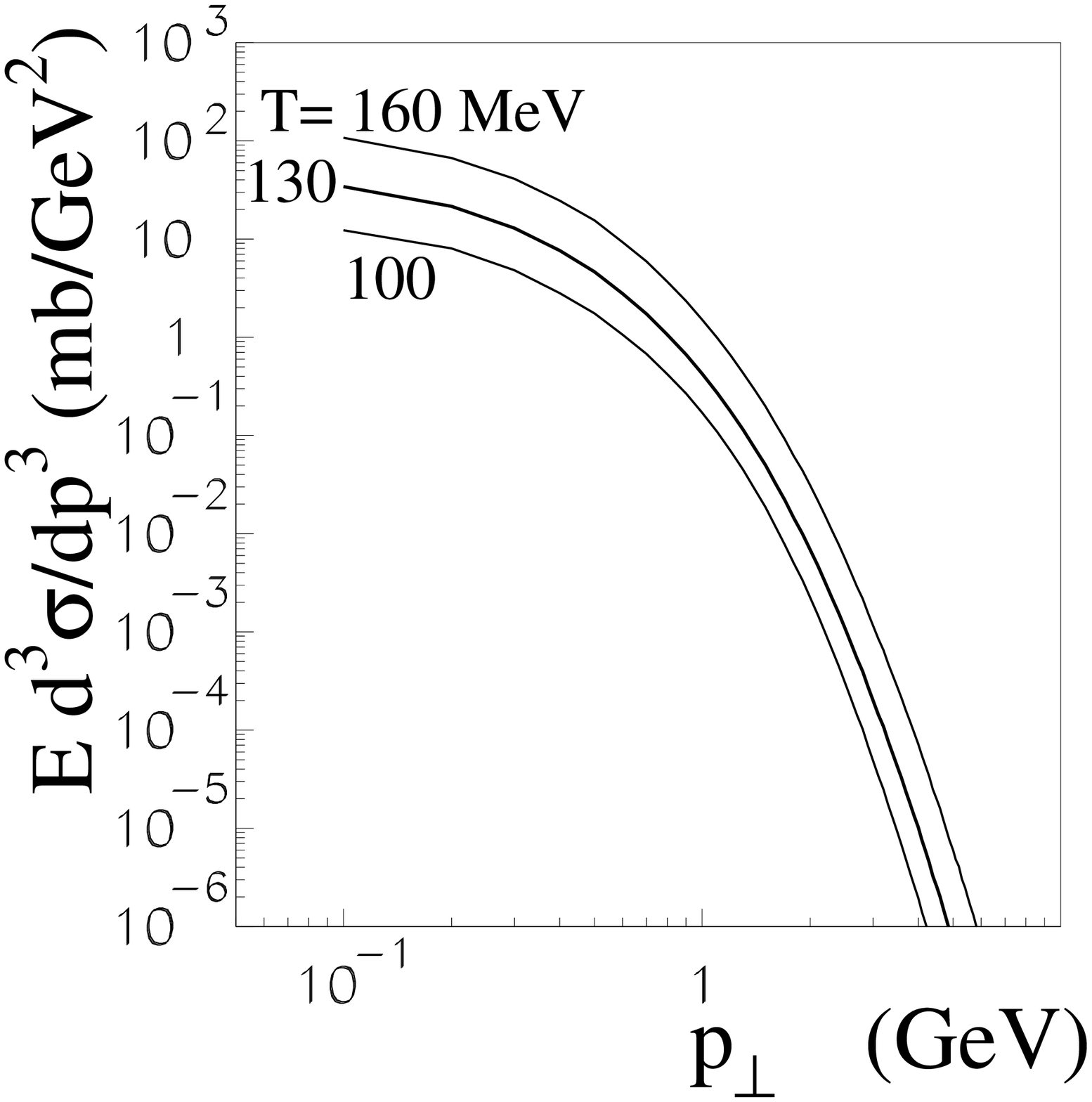}
\includegraphics[width=5.5cm]{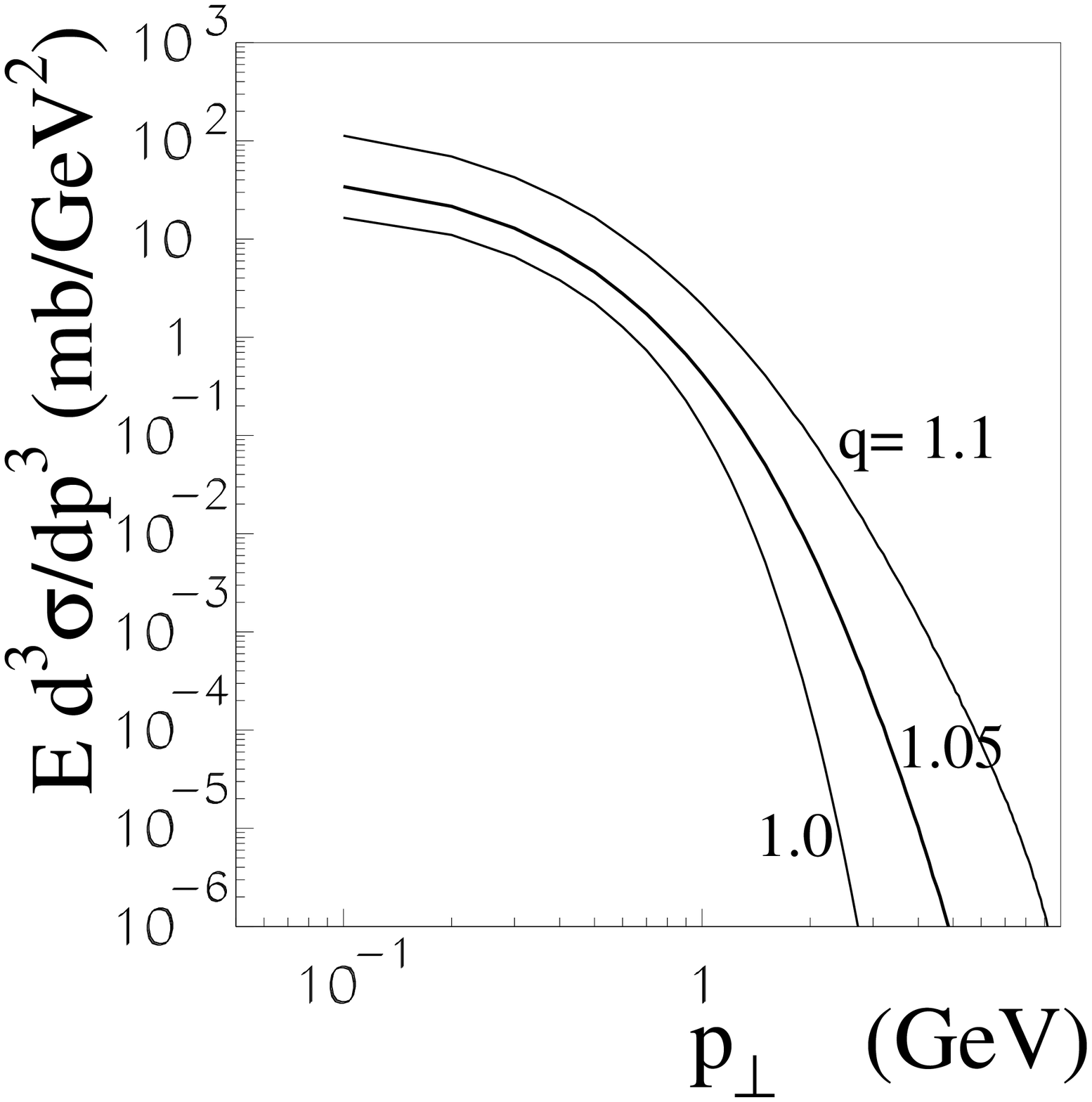}}
\caption{Distributions of $\pt$ for charged long-lived hadrons calculated for different
$V$, $T$, and $q$ values around $T$=130 MeV, $V$=20 fm$^3$ and $q$=1.05.
\label{qtv}}
\end{figure}

Examples of final transverse momentum distributions are shown in Fig.~\ref{qtv} for
different thermodynamical parameters: $T$, $V$, and $q$.
Closer inspection of the $\pt$ distributions
gives reasons to reduce the number of (independent) parameters which should be used when
comparing model calculations with the measured data in the high $\pt$ region. 
The dependence on the hadronization volume $V$ for not very small $\pt$, as seen in the 
left graph in Fig.~\ref{qtv}, is not crucial. 
Additionally, the parameter $T$
(hereafter it 
will be called `temperature', bearing in mind 
that in the non-extensive thermodynamics, for $q > 1$, 
its meaning is not so obvious)
and $V$ are strongly correlated.
In Ref.~\cite{Becattini:1997rv}, for the fitting procedure, the parameter $VT^3$ has been 
chosen instead of $V$.
The main subject of the present paper is not the multiplicity, but rather the shape of the
transverse momentum distribution (and mainly for high transverse momenta), thus
the normalization in the $\pt$ region of interest is quite natural to be used. 
The temperature influences mainly relatively small
momenta, not much higher than its actual value, if, of course, the normalization is treated separately.
The general normalization factor
combines both $T$ and $V$ dependencies and,
as detailed calculations confirm, 
when analyzing high $\pt$ data (about and above $\sim$1 GeV/c),
the parameter $V$ can be successfully replaced by the overall normalization parameter
with only a slight change in two remaining parameters: $T$ and $q$.

The non-extensivity parameter $q$, the crucial one for the present work, 
determines the asymptotic index of the high $\pt$ 
distribution tails. The following approximate formula found in \cite{Beck:2000nz} illustrates this 
\begin{eqnarray}
f(\pt)~\sim~
 \pt
\int \limits_{0}^{\infty} {\rm d}\pl\:
\left[ 1+(1-q)/T\:\sqrt{\pl^2+\pt^2+m^2} \right]^{-q/(q-1)}\sim
~~~~~~~~~~~~~~~~~~~~~~~~~
\nonumber \\
\sim
(\pt/T)^{3/2}
\left[ 1+\pt (q-1)/T \right]^{-\frac{q}{q-1}+\frac{1}{2}}~.
\label{nenept}
\end{eqnarray}
We will compare it with our exact calculation results.

Investigation of the energy dependence of $q$ should give us the answer to the question as to whether
the modified
thermodynamical model can be reasonably applied as a phenomenological description of 
complex QCD principles ``at work'' in the hadronization process.

\section{Application to the data}
The non-extensive framework was successfully applied to transverse momentum data for 
$e^+e^-$ $\rightarrow$ hadrons in \cite{Bediaga:1999hv}. In the present work we present results concerning
$pp$ (and $p \bar p$) reactions. The available data covered quite a wide range of particle interaction
energies. We have started our analysis at $p_{\rm lab}~=~100$ GeV/c
\cite{Ward:1979ig} where the $\pt$ distributions match quite nicely
the exponential behaviour and then through \cite{Adamus:1988xc}
and ISR energies \cite{Alper:1975jm,Drijard:1982zk}, SPS \cite{Albajar:1990an} and \cite{Banner:1985wh} 
up to two Tevatron
energies: $\sqrt{s}=630$ and $1800$ GeV \cite{Bocquet:1996jq,Abe:1988yu}.

Our model parameters (temperature $T$ and non-extensivity $q$) have been adjusted (for $\pt>0.5$ GeV)
to describe measured invariant $\pt$ distributions. The absence of any systematic change 
of temperature was found. Thus we set
the value of $T$ equal to 130 MeV and performed the minimization procedures again. 

The first important point we have to mention here is that the reproduction of the data is very good. 

The next finding is a systematic increase of the non-extensivity 
parameter starting from the value of 1 
(i.e., exponential, e.g.,  classical, $\pt$ distribution tails) at $\sqrt{s}\approx 10$ GeV.

The near-perfect match as obtained suggests a check to see if there exists any evidence at all, 
that the non-extensivity parameter
may have a non-unique value for a fixed interaction energy. Some interesting statements,
suggesting such $q$ behaviour (but for the longitudinal phase space) 
have been published recently \cite{Navarra:2003am}. For example, it is possible that
the non-extensivity varies with the actual multiplicity of created particles, if so it 
could be related somehow to, for example, the 
impact parameter of the colliding hadrons \cite{Wibig:2001ix}.

The spread of $q$ was allowed to be Gaussian 
(with the mean value and dispersion as free parameters 
to be adjusted). Higher degree polynomials have also been 
tested, but no improvement was found.

The fitting procedure was repeated again for all data sets listed above with the $T$ parameter
released free again at the first step. 
No important energy dependence was noticed here either, so we fixed it again at 130 MeV.

The final transverse momentum distributions are presented in Fig.~\ref{finalpt1}

\begin{figure}
\centerline{
\includegraphics[width=7.cm]{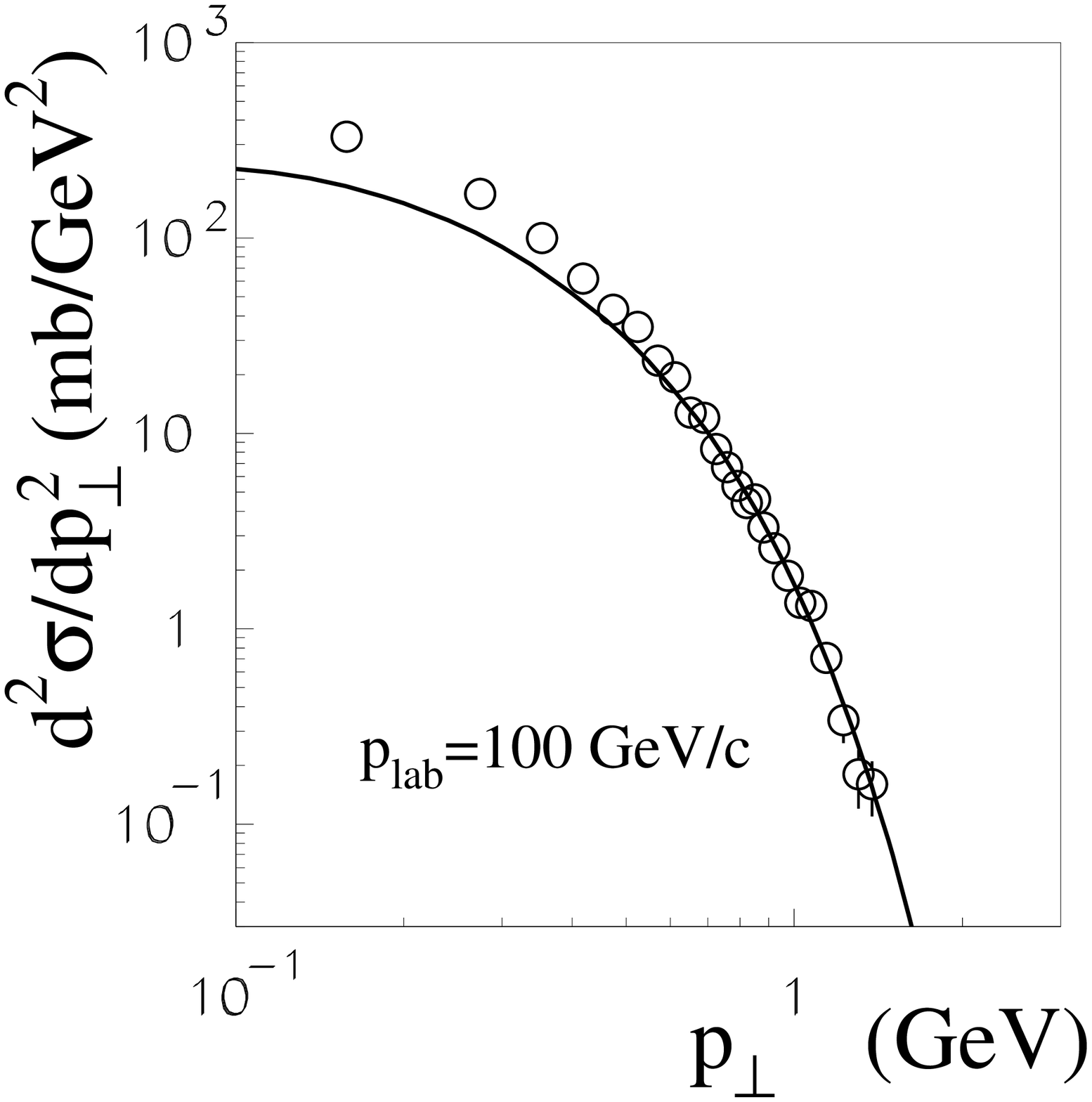}
\includegraphics[width=7.cm]{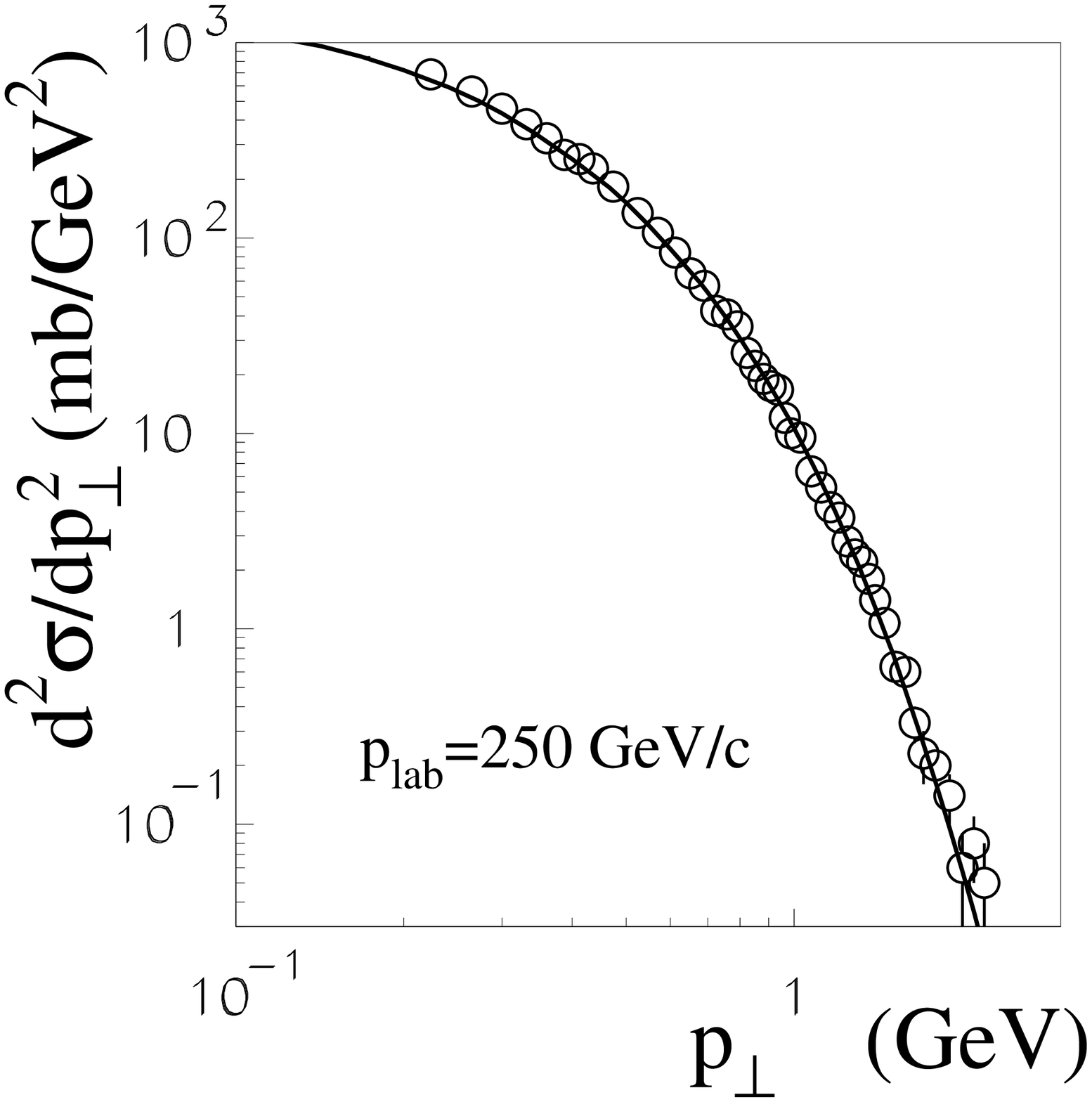}}
\centerline{
\includegraphics[width=7.cm]{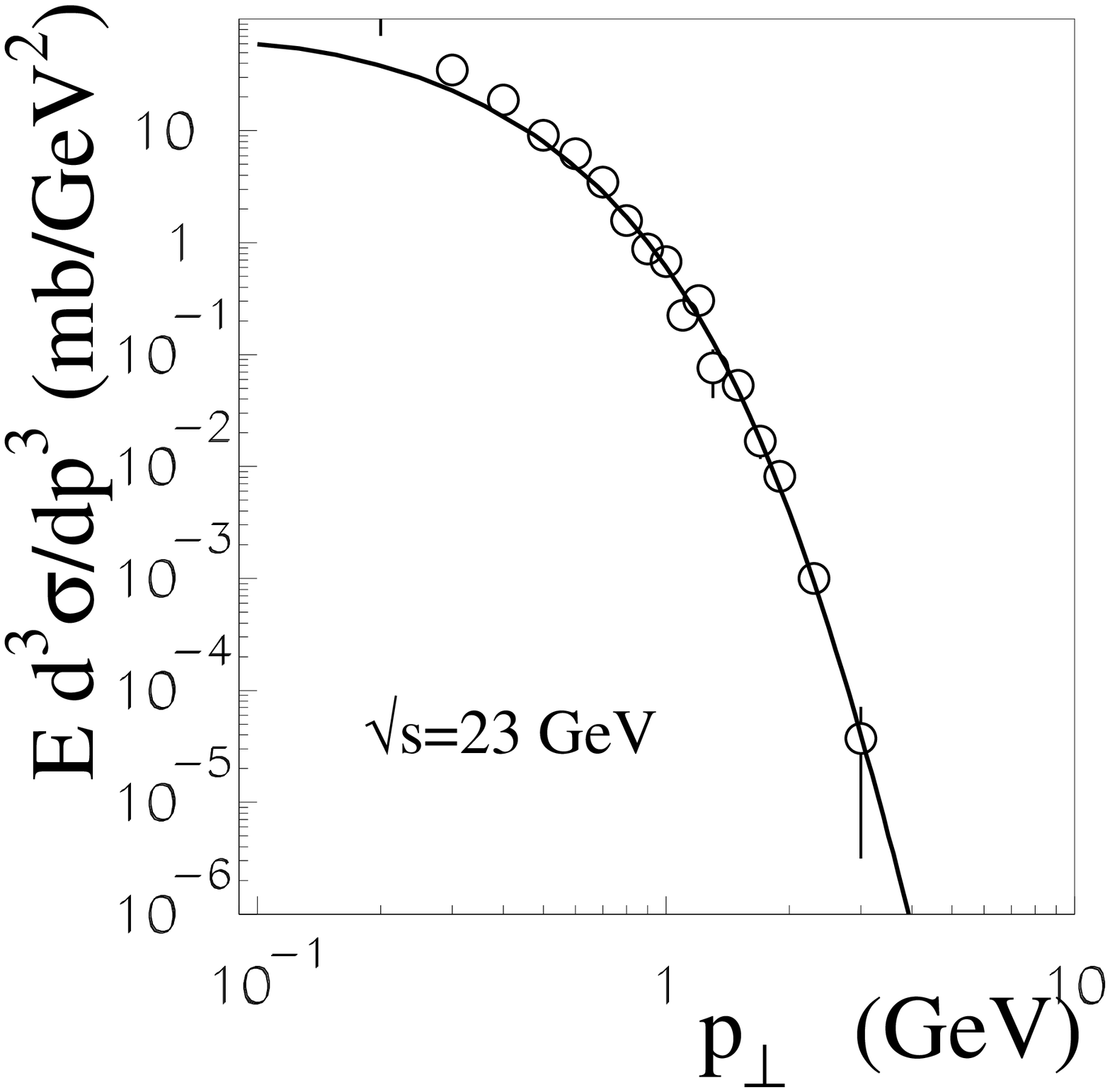}
\includegraphics[width=7.cm]{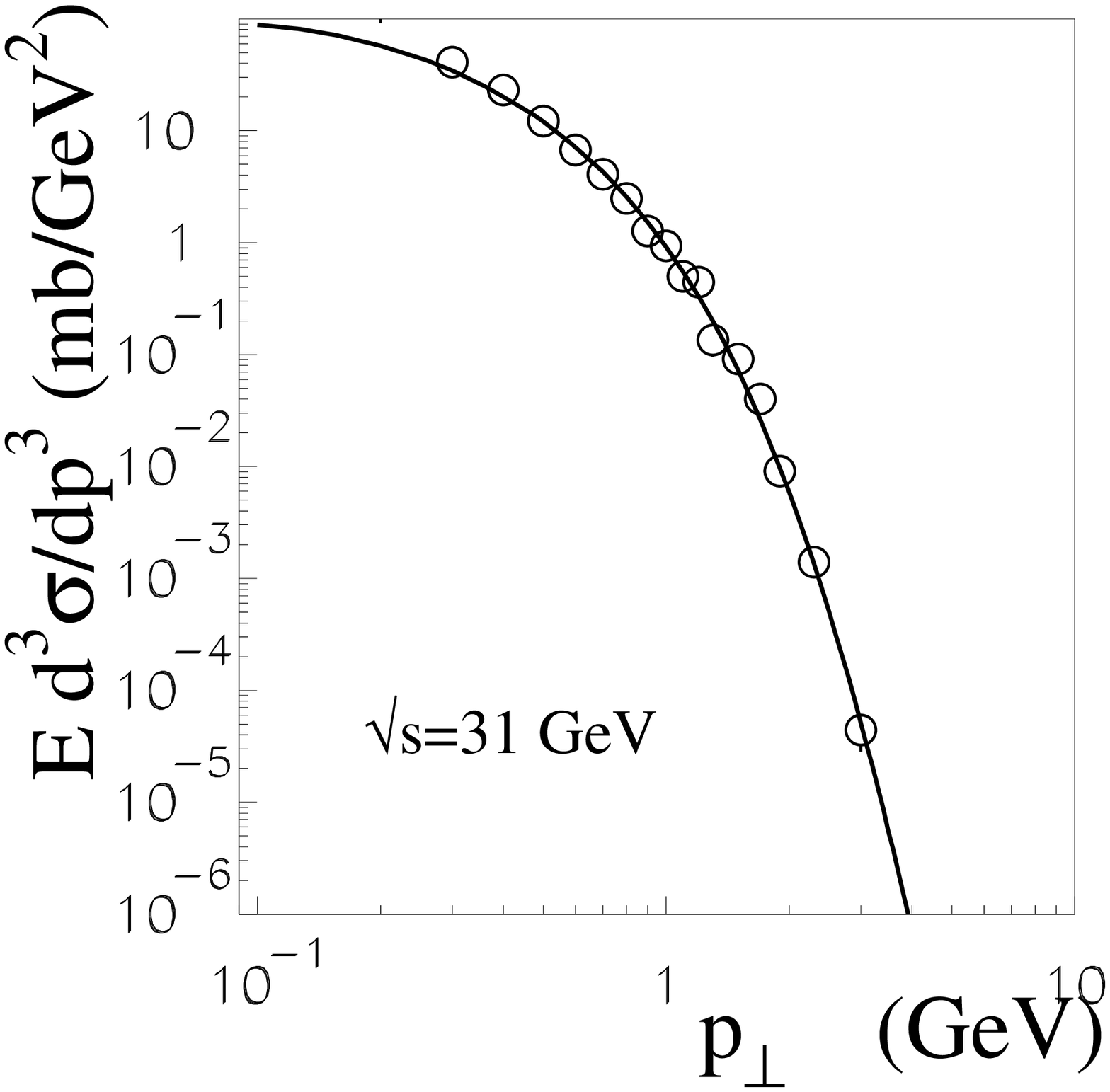}}
\centerline{
\includegraphics[width=7.cm]{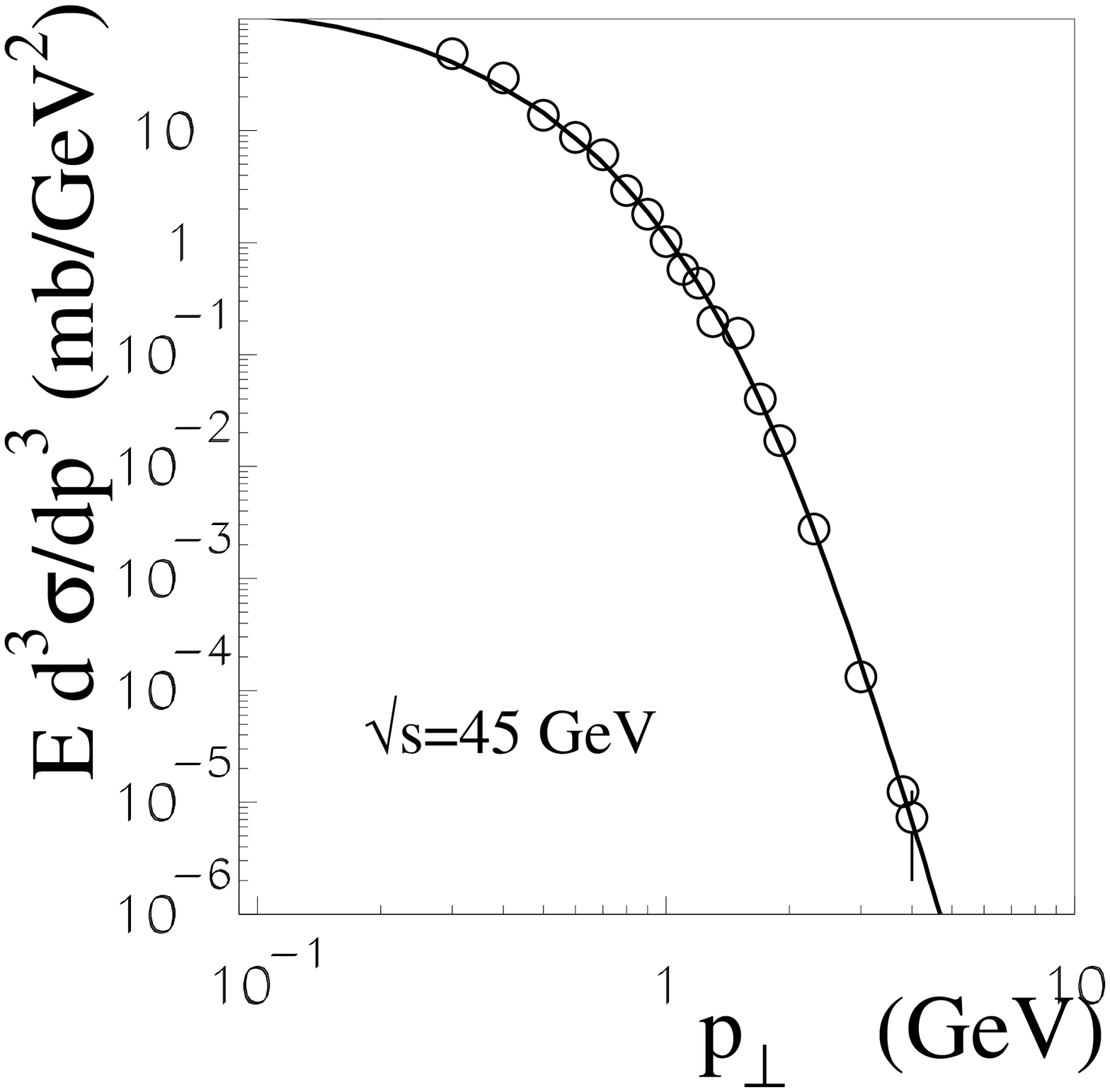}
\includegraphics[width=7.cm]{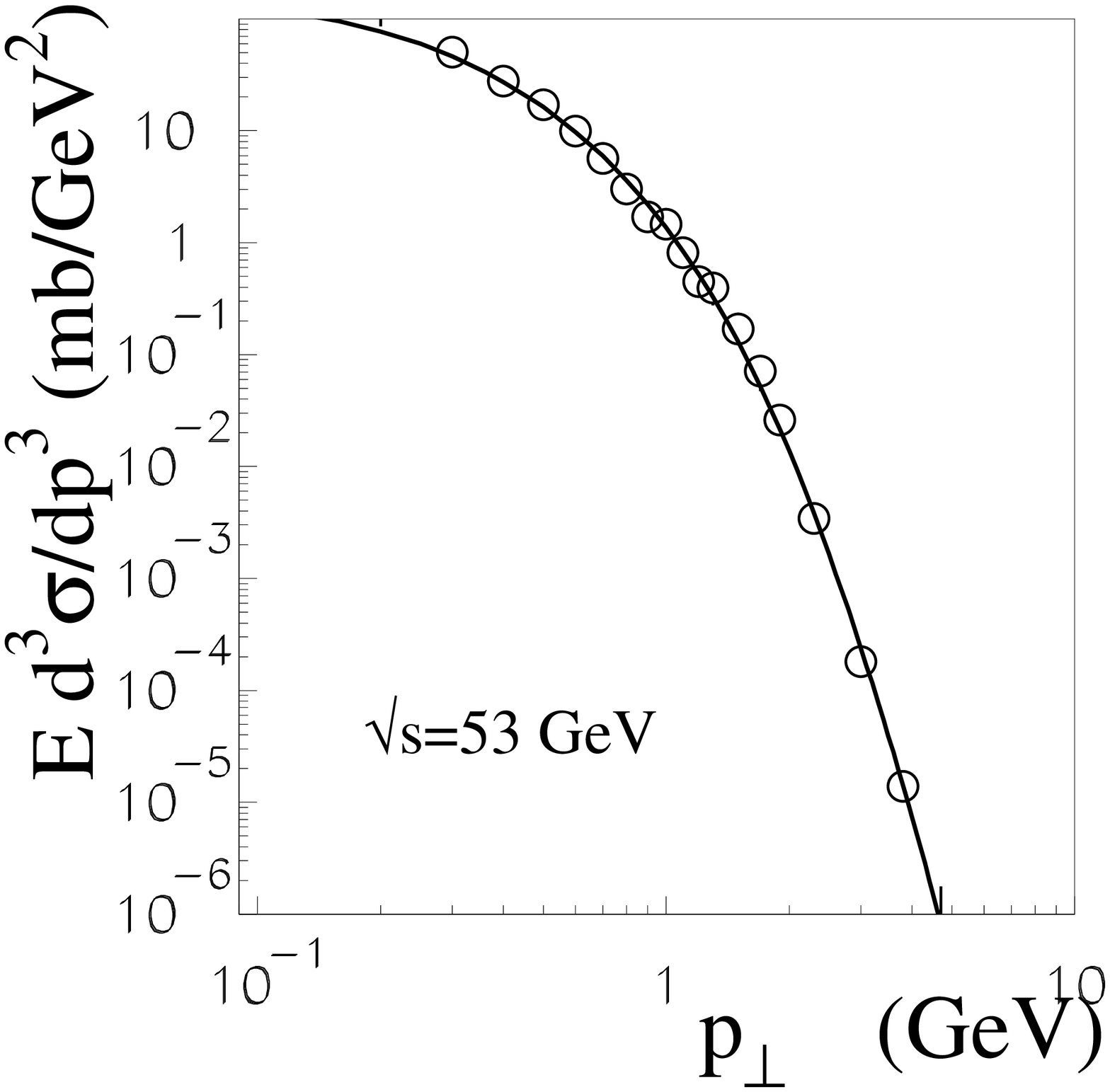}}
\caption{Transverse momentum distributions measured at different energies compared with 
non-extensive fits.
\label{finalpt1}}
\end{figure}
\setcounter{figure}{3}

\begin{figure}
\centerline{
\includegraphics[width=7.cm]{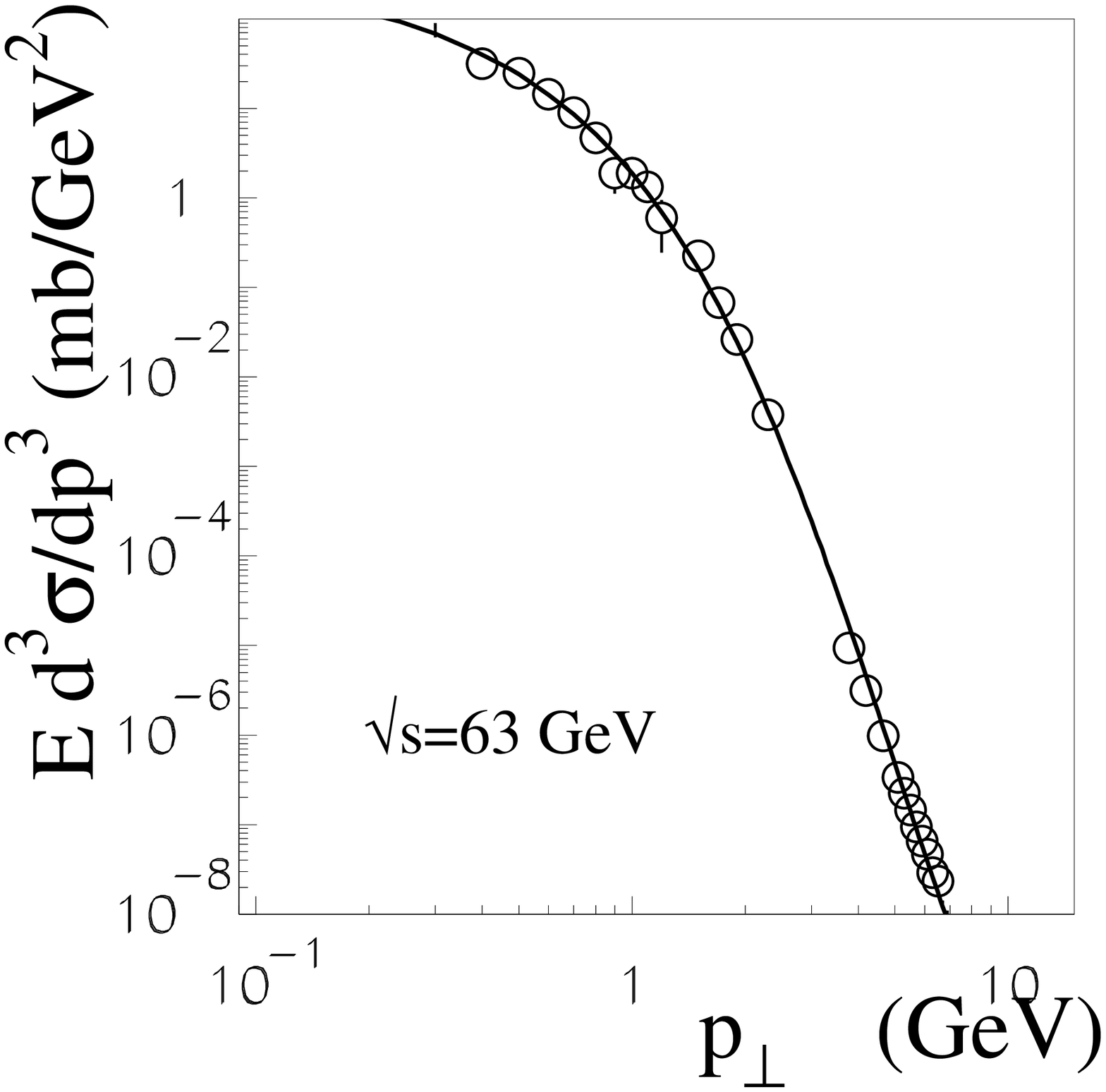}
\includegraphics[width=7.cm]{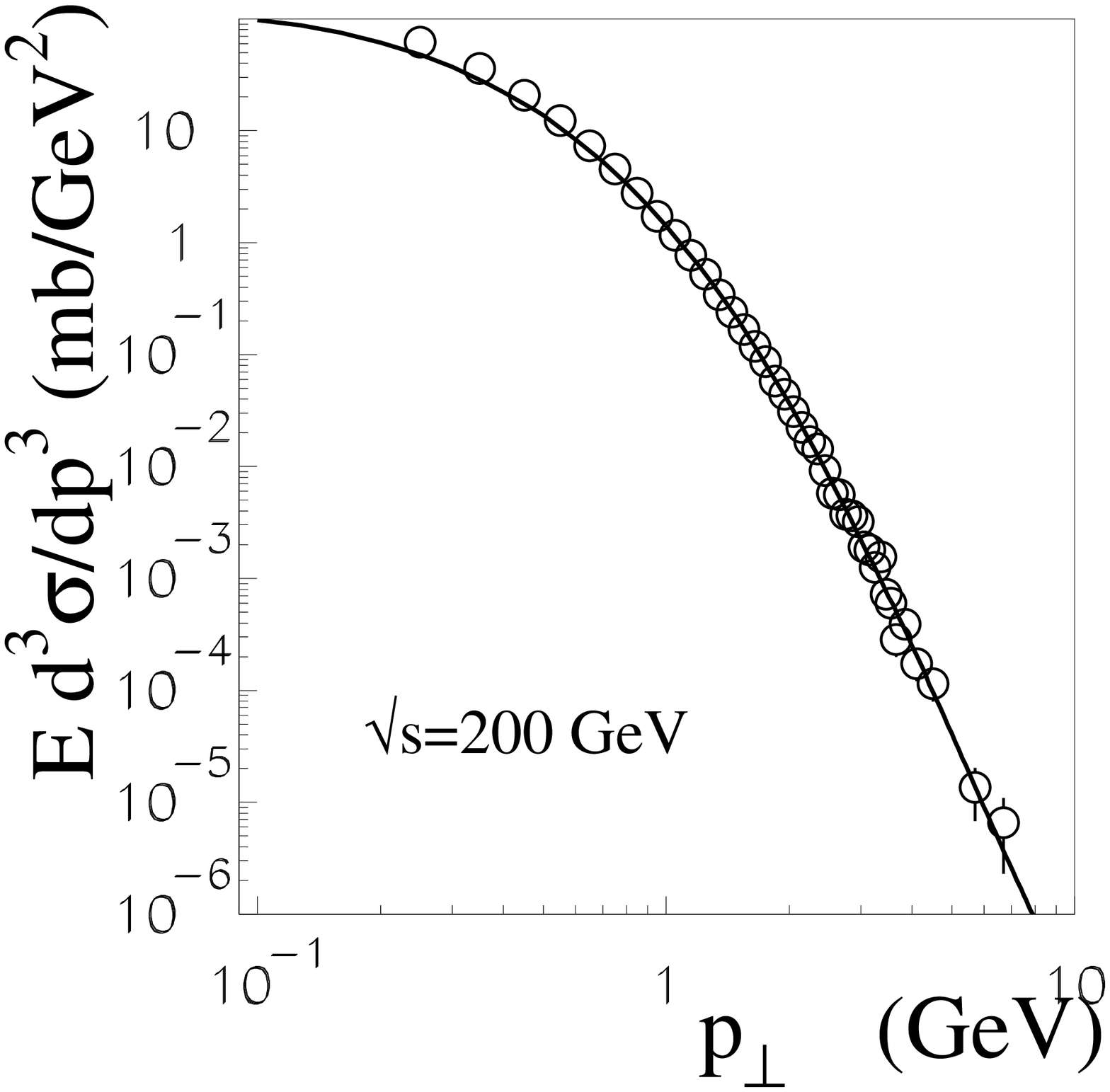}}
\centerline{
\includegraphics[width=7.cm]{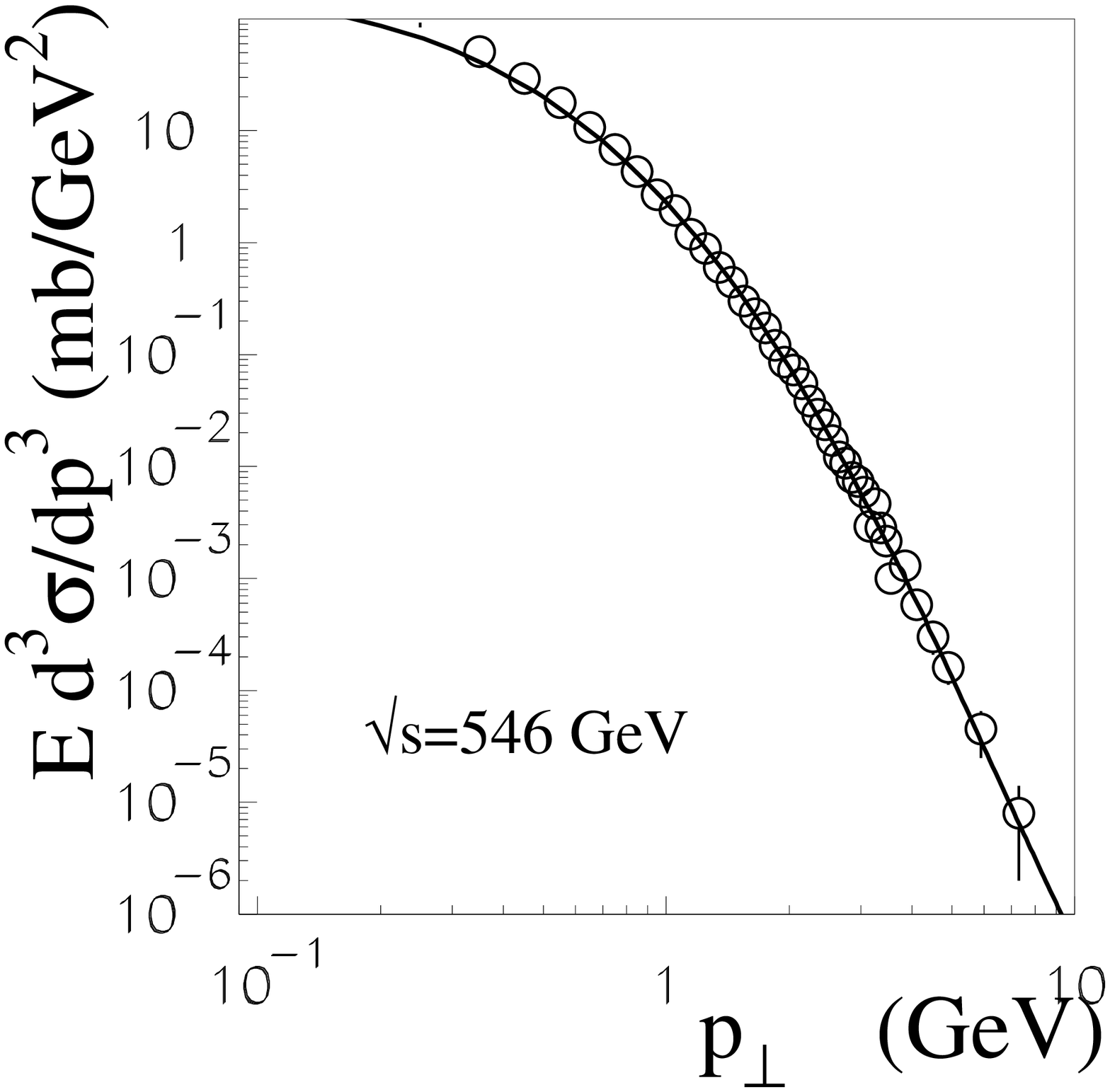}
\includegraphics[width=7.cm]{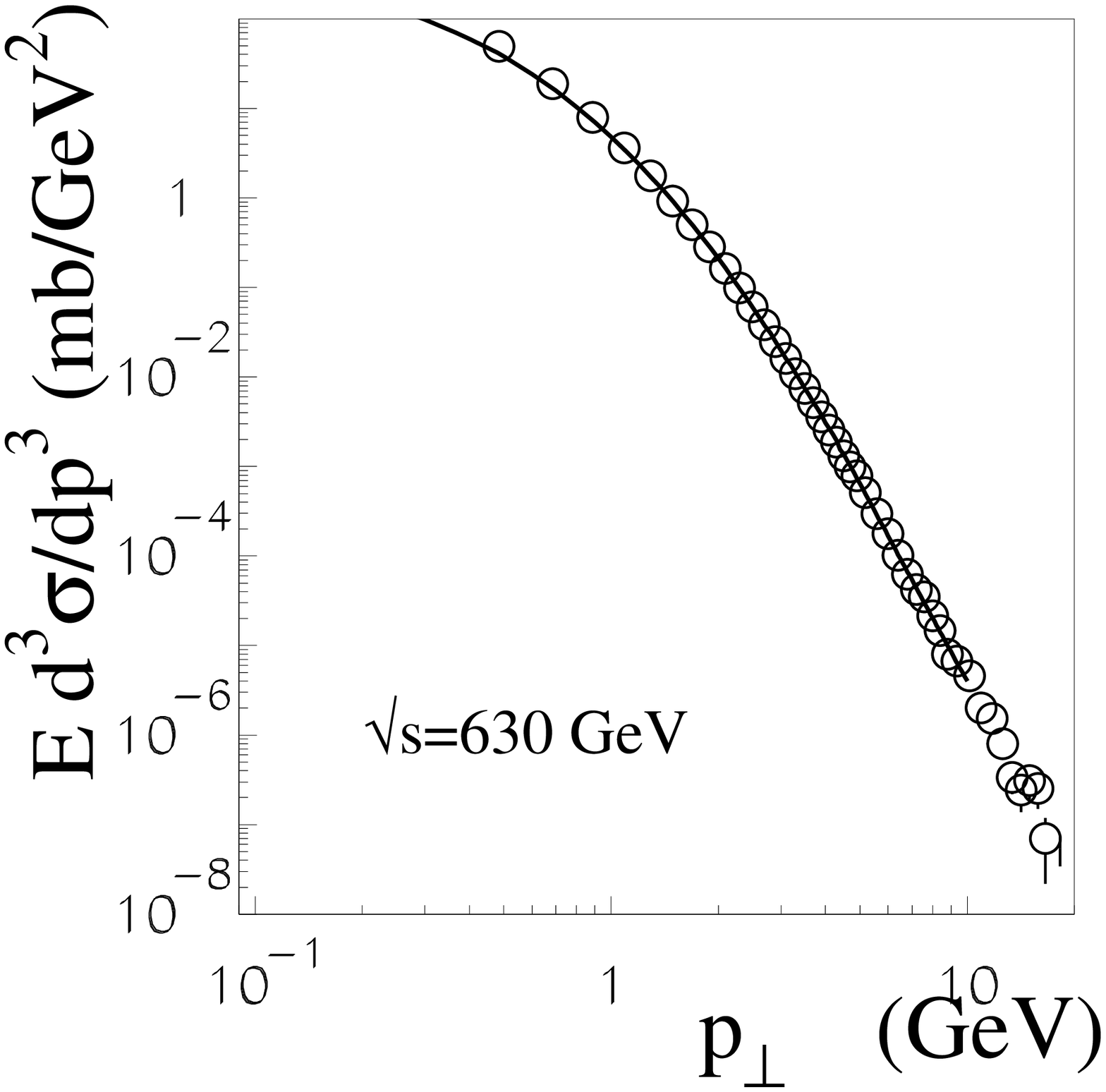}}
\centerline{
\includegraphics[width=7.cm]{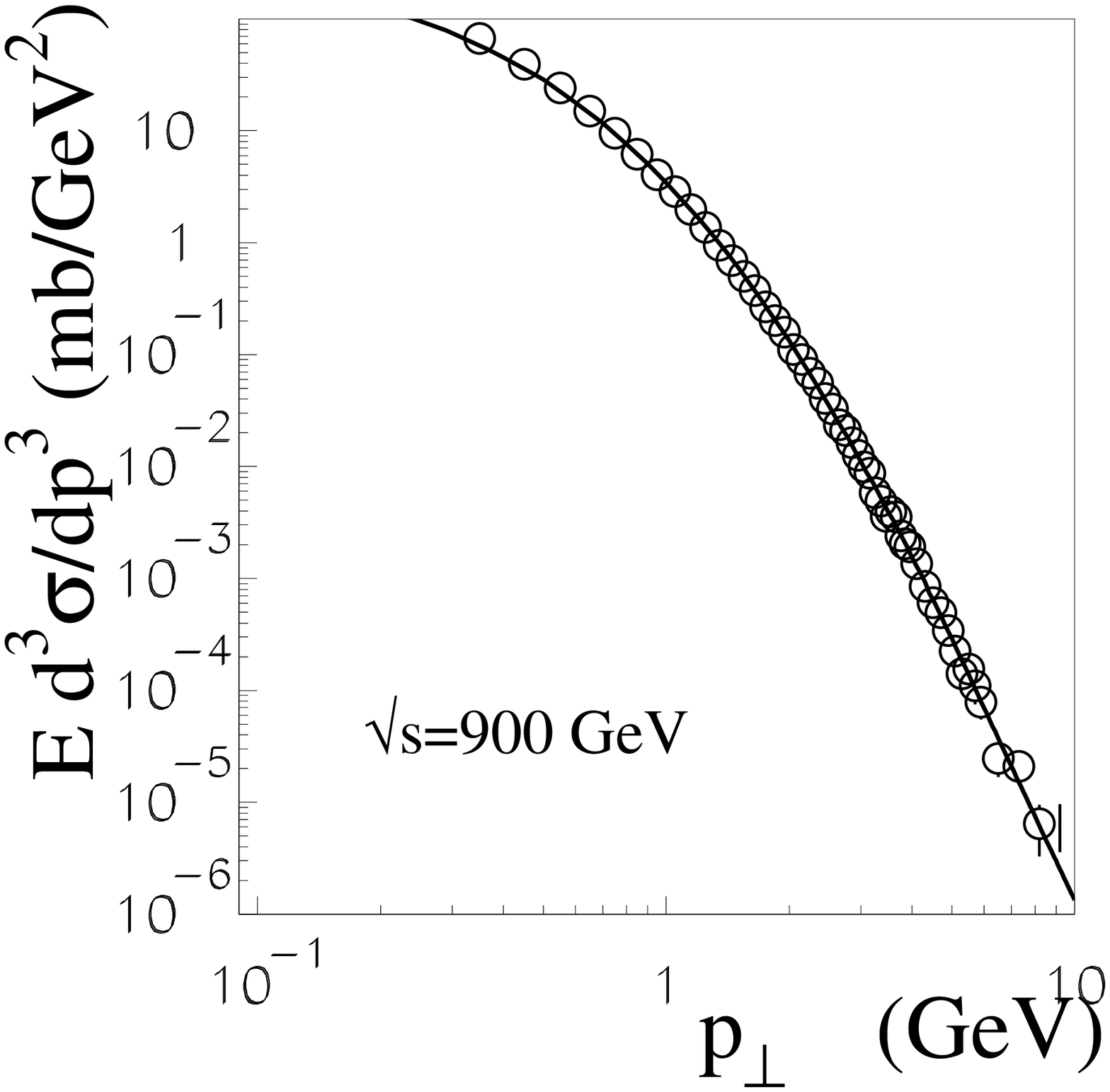}
\includegraphics[width=7.cm]{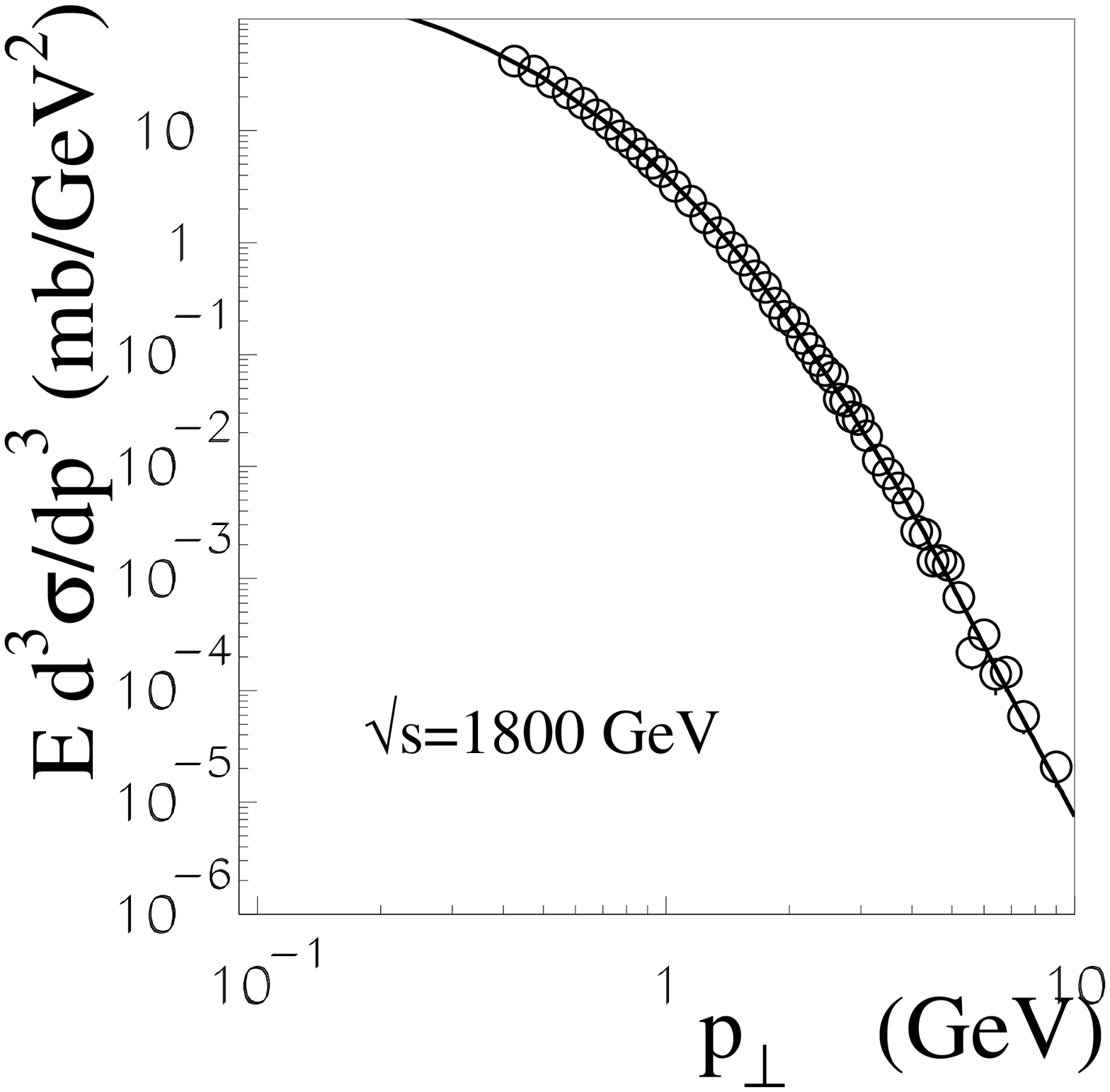}}
\caption{(cont.) Transverse momentum distributions measured at different energies compared with 
non-extensive fits.
\label{finalpt2}}
\end{figure}

\begin{figure}
\centerline{
\includegraphics[width=7.cm]{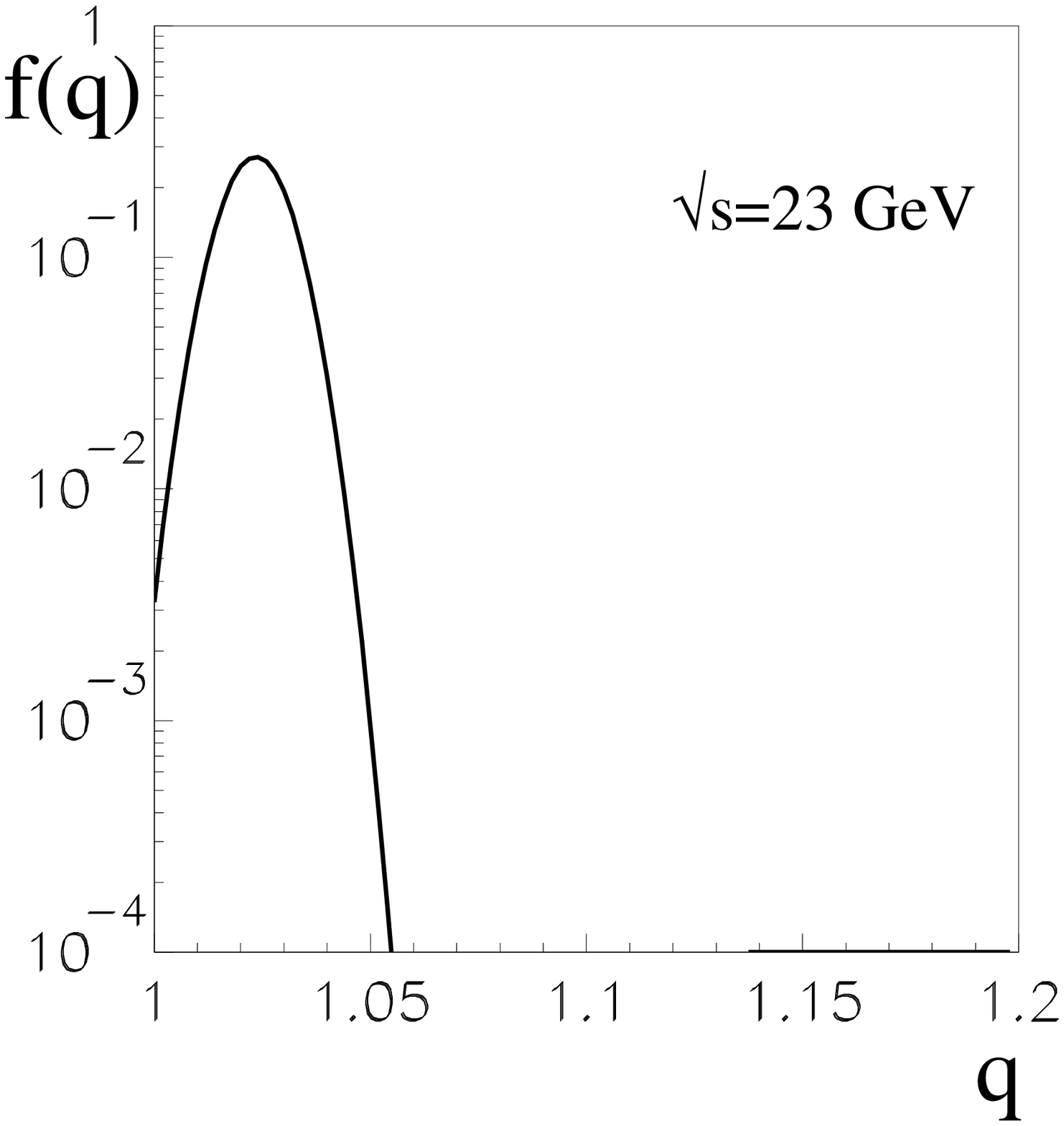}
\includegraphics[width=7.cm]{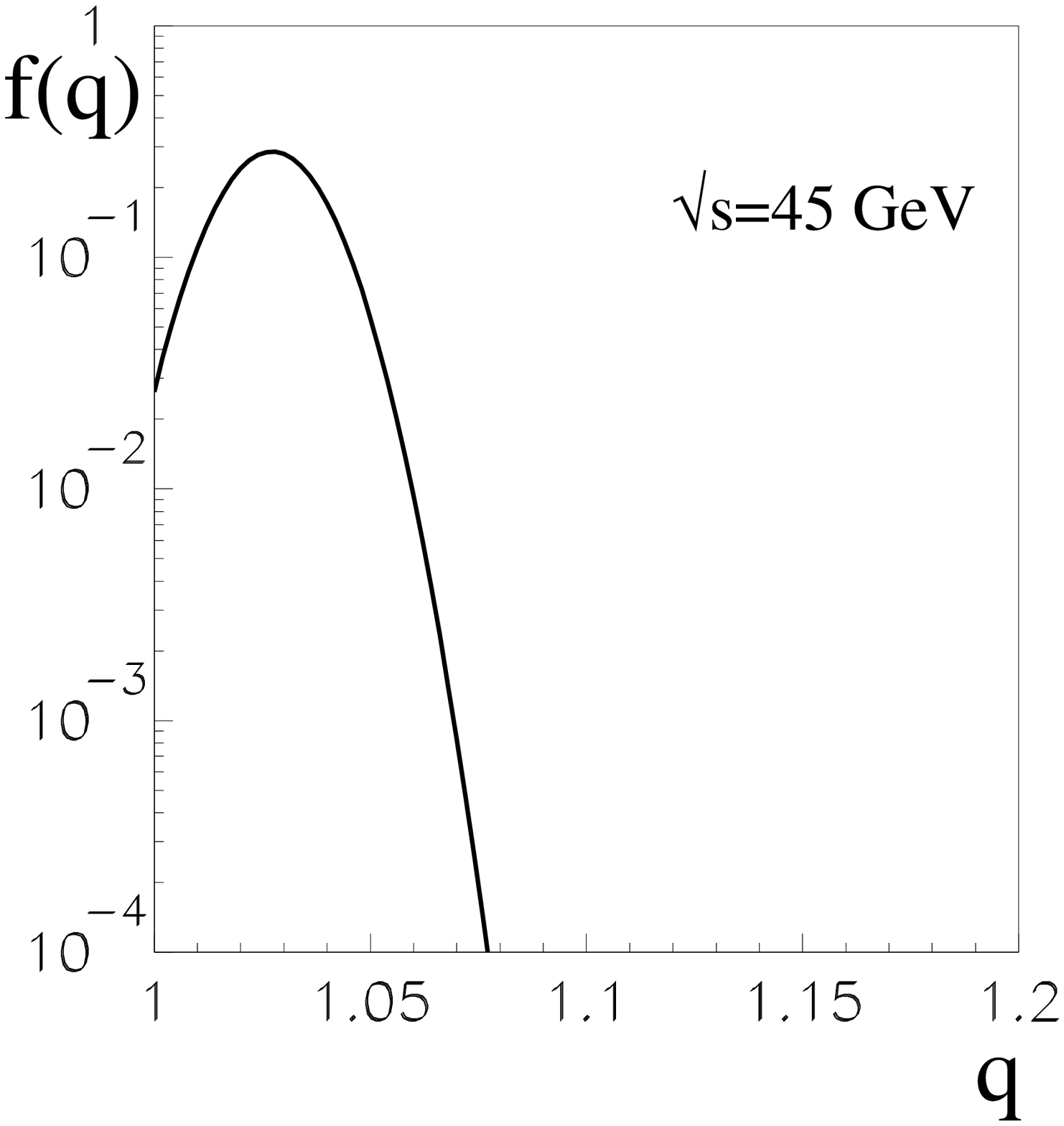}}
\centerline{
\includegraphics[width=7.cm]{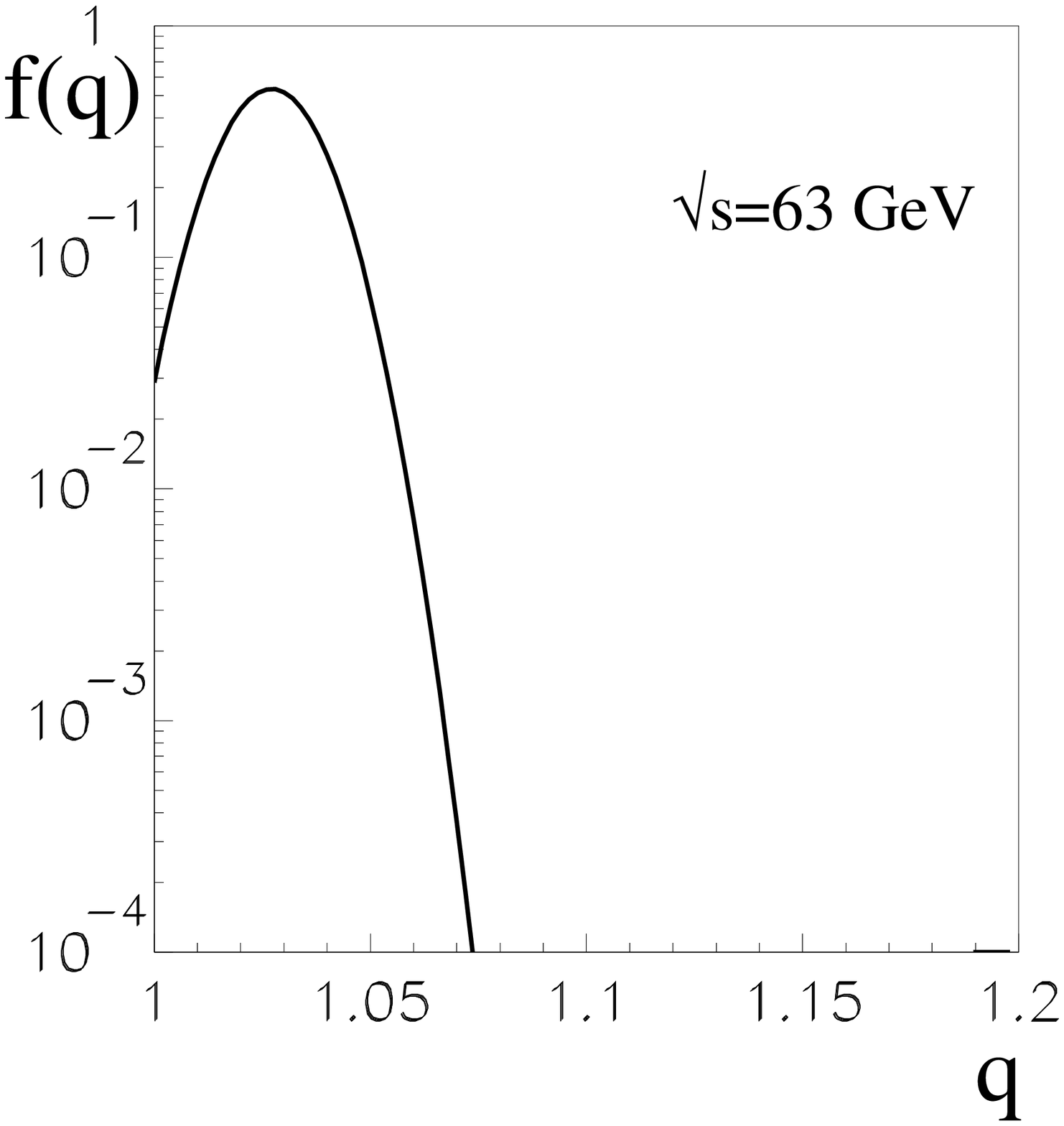}
\includegraphics[width=7.cm]{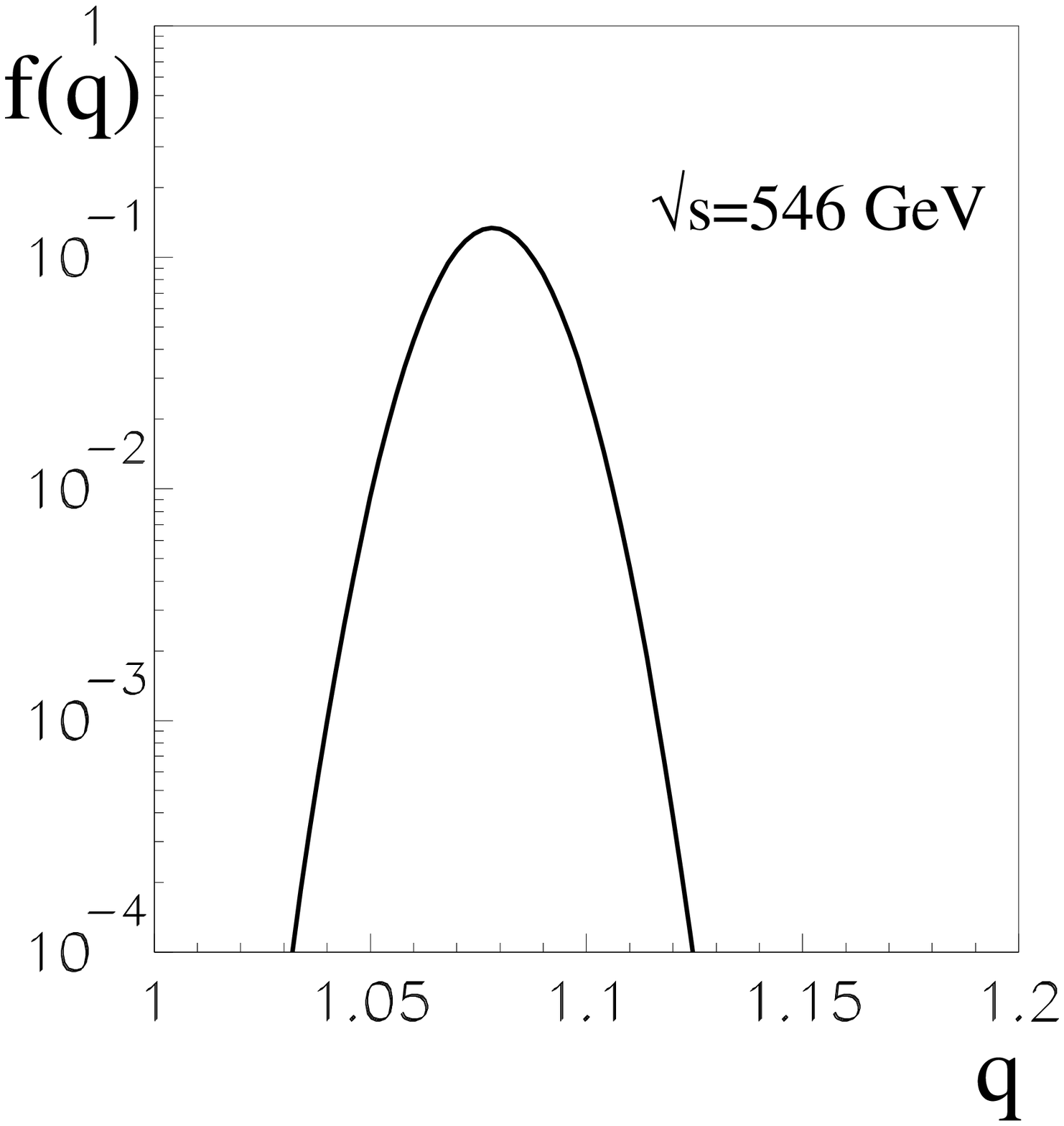}}
\centerline{
\includegraphics[width=7.cm]{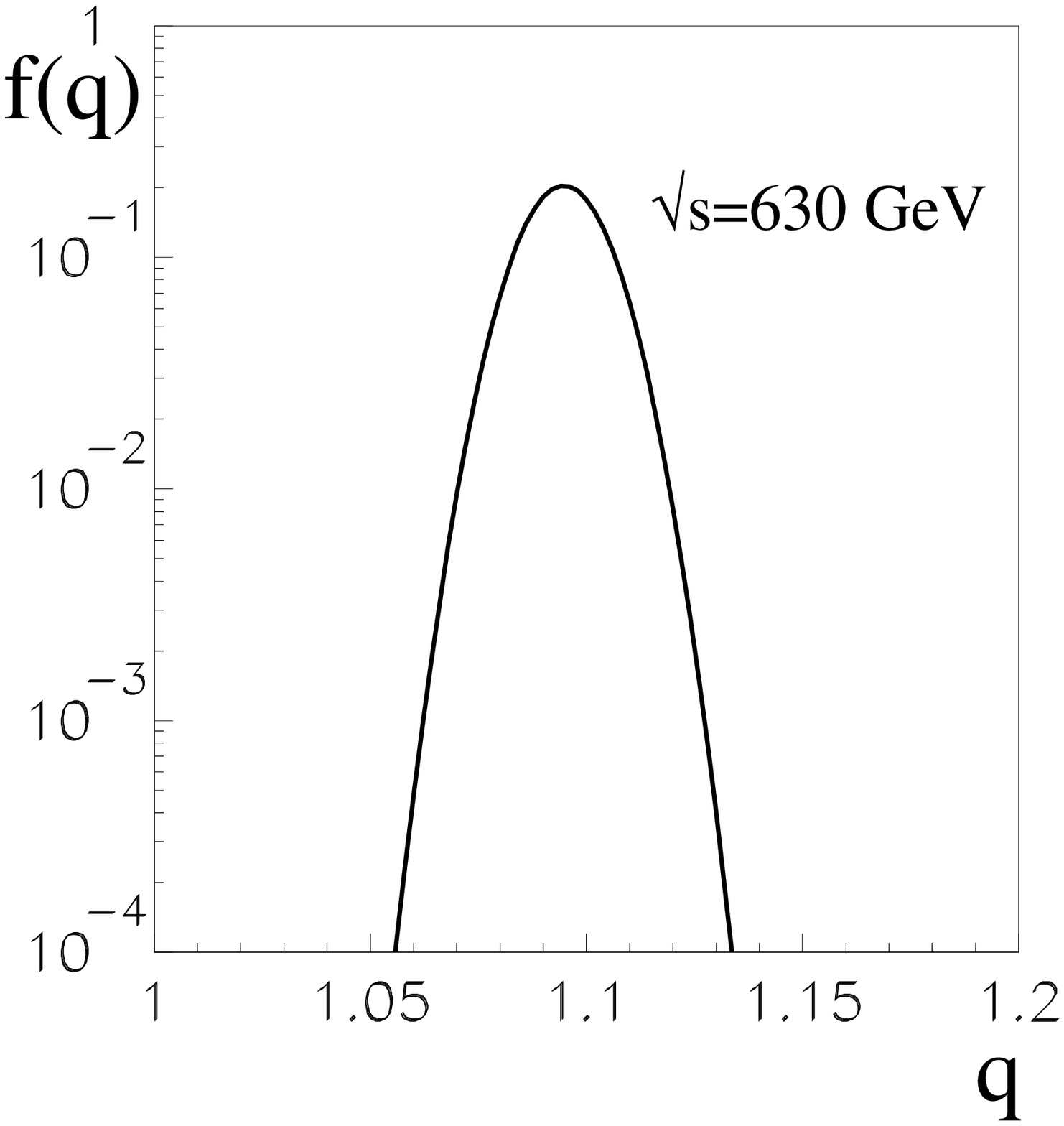}
\includegraphics[width=7.cm]{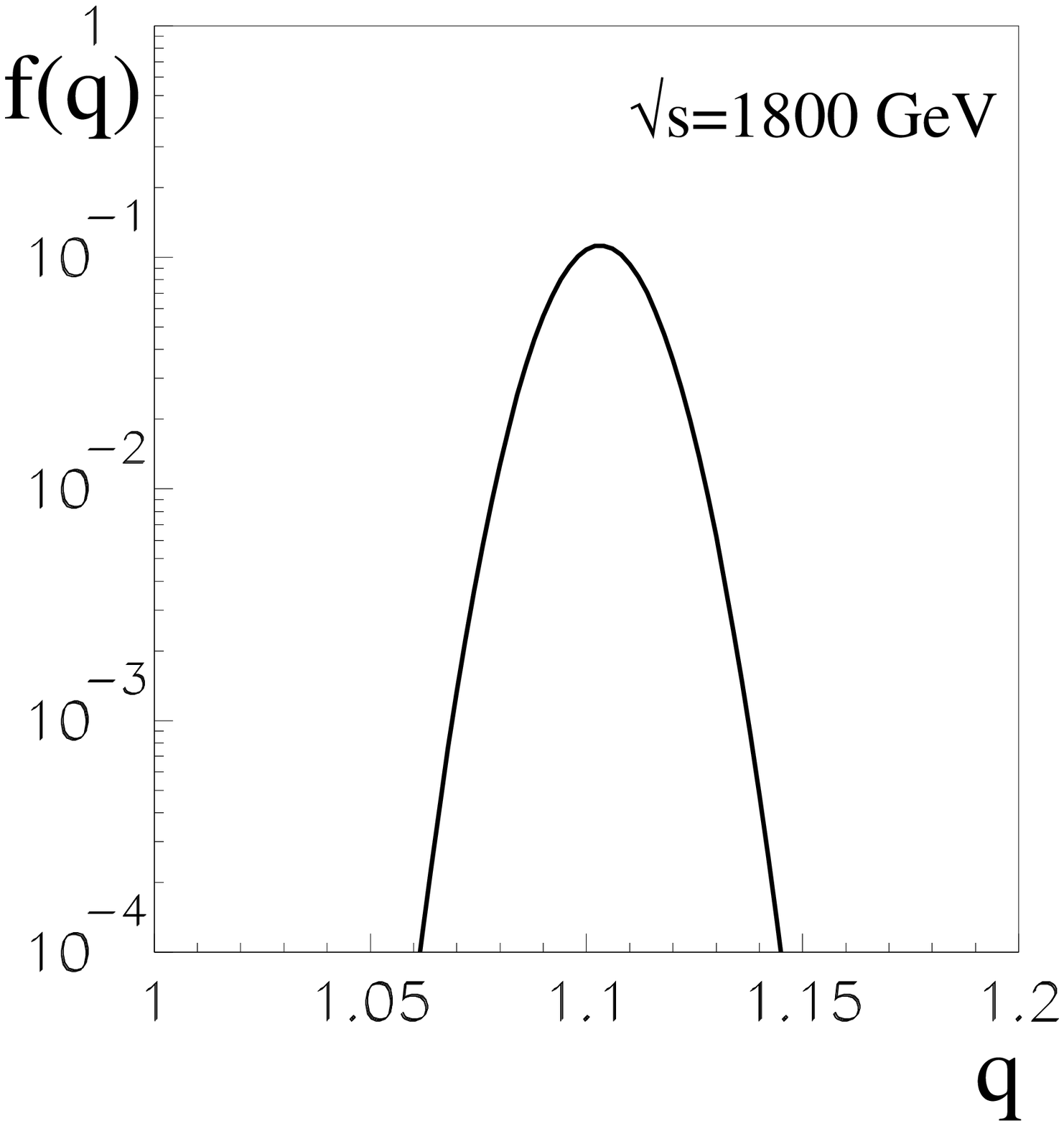}}
\caption{Some examples of distributions of the non-extensivity parameter $q$ giving the best description
of the data shown in Fig.~\ref{finalpt1}.
\label{qdist}}
\end{figure}

The distributions of the $q$ parameter starts from $\delta(q-1)$ for sub-ISR energies, and then,
when the average $q$ differs from 1 they become Gaussian (truncated below 1). 
Some examples are presented in Fig.~\ref{qdist}
In all cases the widths of the $q$ distributions are very small (their dispersions are about $0.01$). This 
allows us to claim that the transverse momentum distributions for a given interaction energy can
be well described by a single value for the non-extensivity parameter.
This means that the
parameter does not depend on the actual multiplicity (impact parameter) or any other
interaction characteristic which may fluctuate from event to event. Its value
is determined only by the available center of mass 
energy. It should be mentioned here, that this statement supports the non-extensive 
thermodynamical treatment of the hadronization process making it simple and clear.

\begin{figure}
\centerline{
\includegraphics[width=8.5cm]{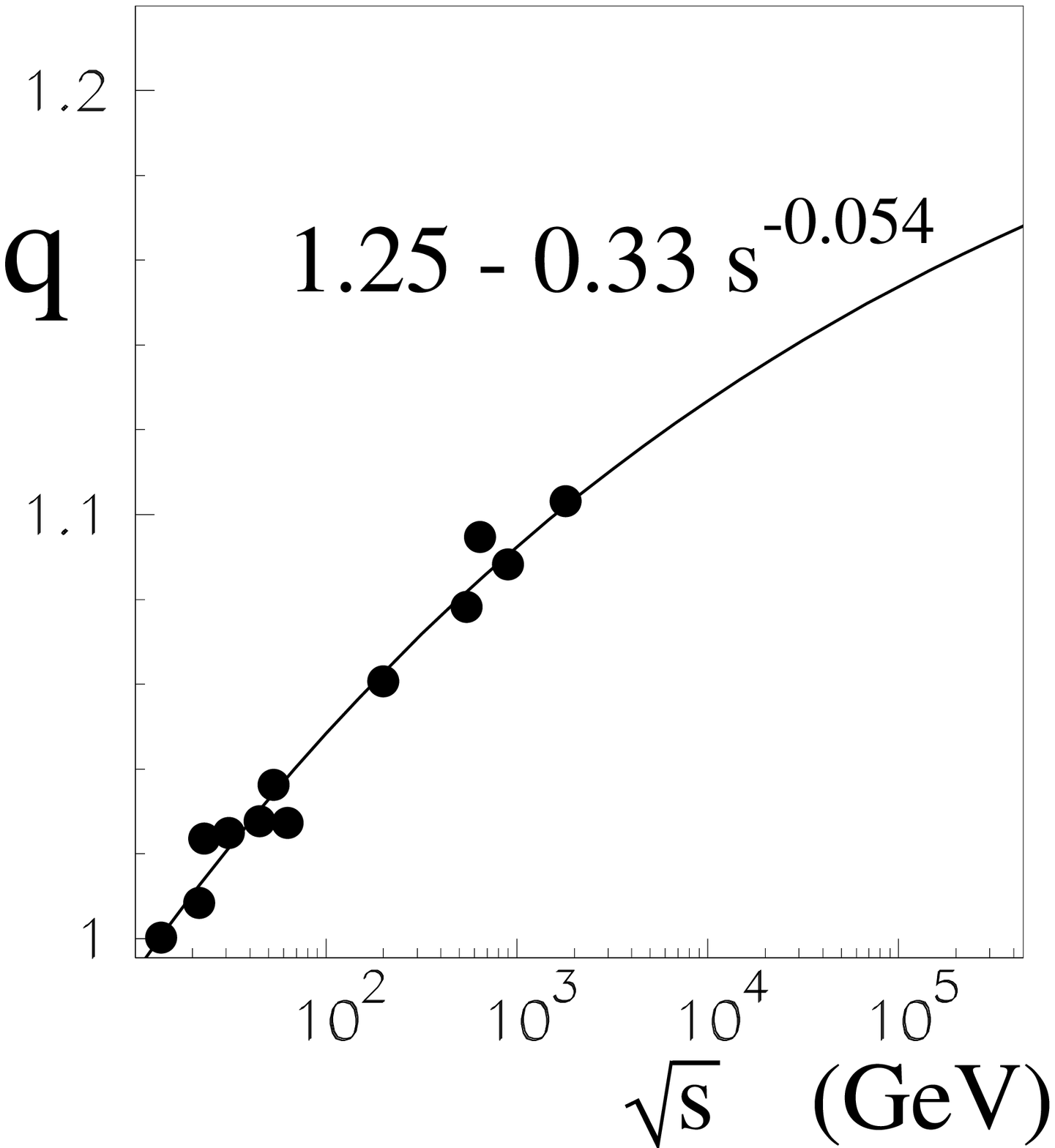}}
\caption{Energy dependence of the average value of the non-extensivity parameter $q$ obtained from the $\pt$
distributions.  The functional fit is shown by the line.
Uncertainties at each point coming from the fitting procedure are much smaller than
plotted points. The spread with respect to the line is a result of differing systematic
uncertainties
in the $\pt$ distributions measured by different experiments.
\label{qs}}
\end{figure}

However, most important is the intriguing regularity as seen in 
the Fig.~\ref{qs} where we show how the non-extensivity
parameter $q$ depends on the interaction energy.

The value of $q$ should not exceed 1.25 \cite{Beck:2001ts}, and it is quite reasonable to 
expect this as an asymptotic limit leading to QCD-inspired $\pt\!\!^{-4}$ distribution 
for large transverse momenta at extremely high energies. The simple dependence is found to be of the form
\begin{equation}
q~=~1.25~-~0.33\:s^{-0.054}
\label{qfit}
\end{equation}
as shown in Fig.~\ref{qs}.

\section{Rise of the average transverse momenta}
\begin{figure}
\centerline{
\includegraphics[width=8.5cm]{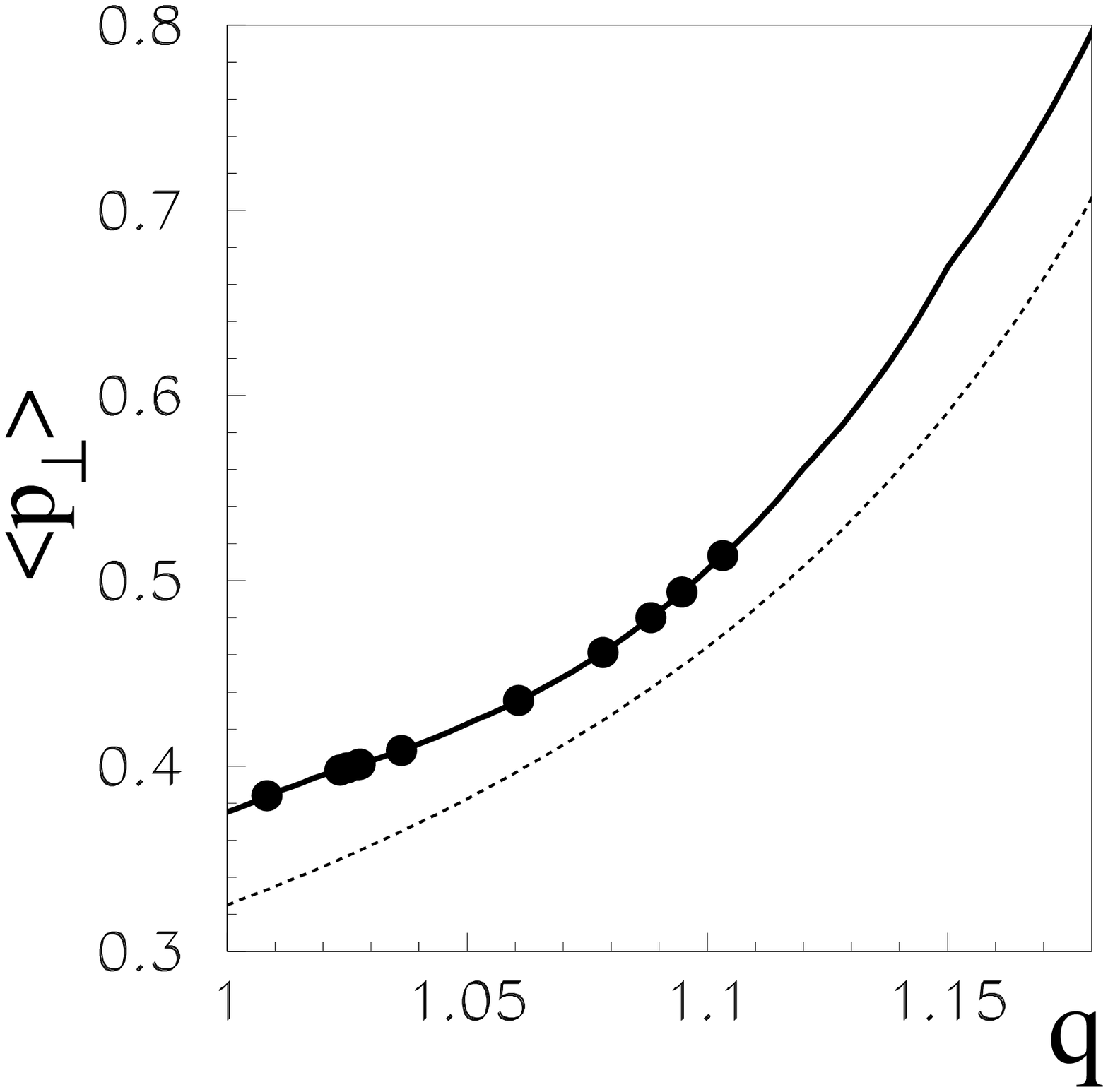}}
\caption{The calculated average transverse momentum of 
long-lived charged hadrons as a function of the non-extensivity parameter $q$.
The dashed line shows the approximation given by Eq.(\ref{avptb}). 
The points represent the mean $\pt$'s calculated from the fits shown in 
Fig.~\ref{finalpt1}.
\label{ptq}}
\end{figure}

The average transverse momentum of particles created in high energy hadronic interactions 
in the non-extensive thermodynamical model increases when the non-extensivity
parameter increases. In \cite{Beck:2000nz} the average of the distribution given by 
Eq.(\ref{nenept}) was calculated.
The form of the dependence is
\begin{equation}
\left\langle \pt \right\rangle ~=~ T\:\frac{1}{2}{5 \over 4\:-\:3q}~.
\label{avptb}
\end{equation}
This is, however, an approximate result (as is Eq.(\ref{nenept})). The
approximation is quite good but it can only give the average value for primary created hadrons. 
We have obtained the average $\pt$ exactly and 
corrected for all the effects of decays of short-lived resonances. It is presented in Fig.~\ref{ptq}
where Eq.(\ref{avptb}) is compared with the results obtained from the fits presented
in Fig.~\ref{finalpt1}.

\begin{figure}
\centerline{
\includegraphics[width=8.5cm]{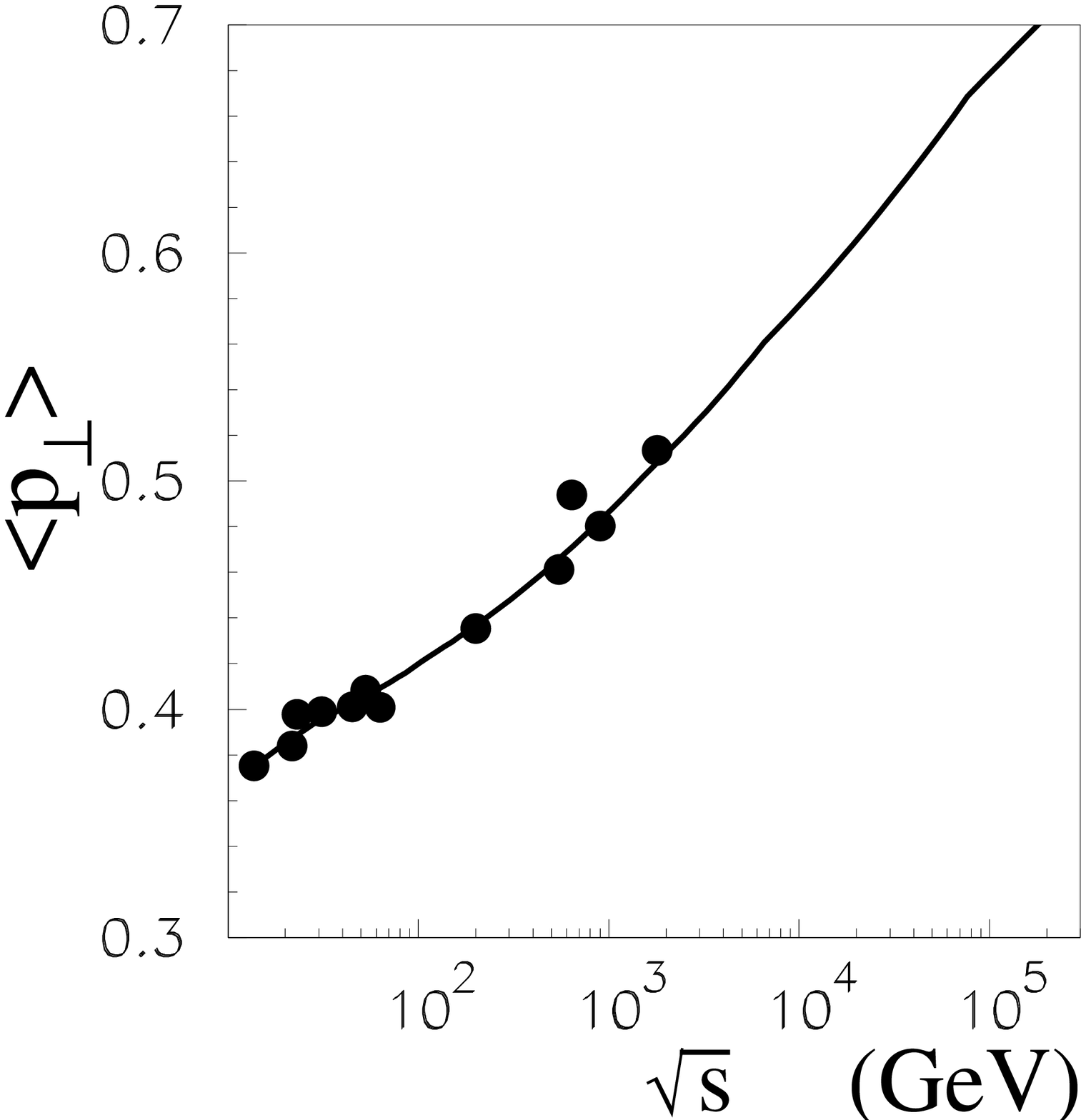}}
\caption{The average transverse momentum of 
long-lived charged hadrons as a function of the available interaction center of mass energy 
for our model with 
$q(s)$ given by Eq.(\ref{qfit}). The points are as in Fig.~\ref{ptq}.
\label{pts}}
\end{figure}

Using Eq.(\ref{qfit}) we can present the average $\pt$ as a function of energy; the result is 
shown in Fig.~\ref{pts}. 

The existence of the 1.25 asymptotic limit for $q$ and the clear and well defined 
rise of $q$ observed in the data allows us to extrapolate the results to much higher 
energies with a high degree of confidence. The line plotted in Fig.~\ref{pts} shows a
moderate $\langle \pt \rangle$ increase (as $s^{0.037}$ or of about 0.1 GeV per decade
of the center of mass available energy) for large interaction energies. 

This is relevant for the creation of new Monte Carlo generators and interaction models 
used in cosmic ray physics simulations of extensive air showers (EAS) at very high energies. 
Just now is the moment when the new
generation of giant EAS arrays are built and new data from, e.g., Auger Observatory 
for energies as high as 10$^{20}$~eV are expected \cite{Blumer:2003sm}. 
Their proper interpretation and the accurate
primary particle energy estimation requires detailed knowledge of the expected 
lateral particle distribution which is determined in part by the transverse particle spread
high in the atmosphere where the particles interact.
The energies are, of course, much above those
energy created at the accelerators. 

According to \cite{corsika} the average $\pt$ used in all models implemented in the CORSIKA code, one of
the most widely used EAS simulation programs, at 10$^{19}$eV is 0.55~GeV while our fit
suggests a value which is about 20\% higher.

Cosmic ray experiments have been delivering for a very long time data related to average 
transverse momenta at energies exceeding the contemporary accelerator abilities. 
The most straightforward, at first sight, are the calorimeter experiments
at mountain altitudes. The claim of an abrupt and very substantial
 $\pt$ increase at around $\sqrt{s}=1000$~GeV
made in \cite{plovd} was based on observations of events with large values of the $Er$ product in the Tien-Shan
hadronic calorimeter. The registration of the shower of particles and determination of their energies
$E$ and distance to the shower axis $r$ 
enables a determination of the distribution of product $Er$ 
which is related to the transverse momentum by
\begin{equation}
\pt~=~{r \over h}\:{E \over c}~,
\end{equation}
where $h$ is the actual hadron production height.
However, methodological difficulties in the data interpretation (see, e.g, \cite{Capdevielle:1998my})
do not permit such a strong statement.
In spite of this fact $Er$ is still the simplest and promising way to study the transverse spread of hadrons
created in very high energy collisions.

\begin{figure}
\centerline{
\includegraphics[width=8.5cm]{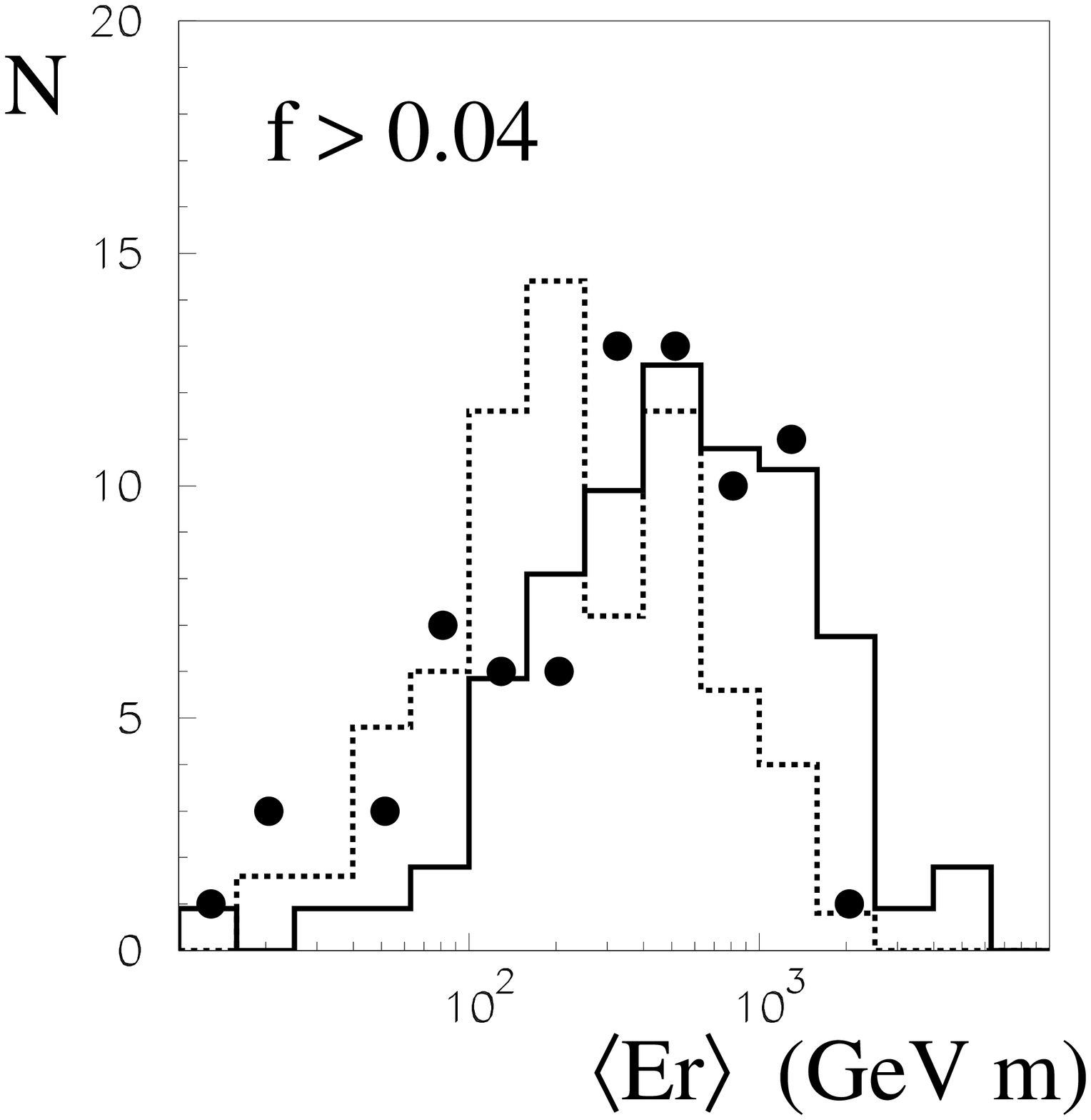}}
\caption{Distribution of $\langle Er\rangle$ in the Chacaltaya-Pamir experiment \cite{pamcha}
compared with predictions
of FRITIOF (dashed histogram) and modified HDPM (solid one) models
\label{er}}
\end{figure}

In \cite{pamcha} the distribution of $Er$ from the joint Chacaltaya-Pamir experiment was published,
again with the conclusion that its shape could not be fully explained by the simulations.
Their data are shown in Fig.~\ref{er} as solid circles and the results of our calculations with two interaction
models are given by the respective histograms. For the shower simulation we used the structure of the 
CORSIKA 
program \cite{Heck:1998vt}. The well known FRITIOF \cite{Andersson:1993iq} model was implemented to see
if the event generator widely used in accelerator physics works well also in the cosmic ray domain. The second 
model was the version of the default
CORSIKA model called HDPM (\cite{Capdevielle:1989ht} with further improvements)
in which we change the generation of transverse momenta according to the results obtained in the
present paper.

\begin{figure}
\centerline{
\includegraphics[width=8.5cm]{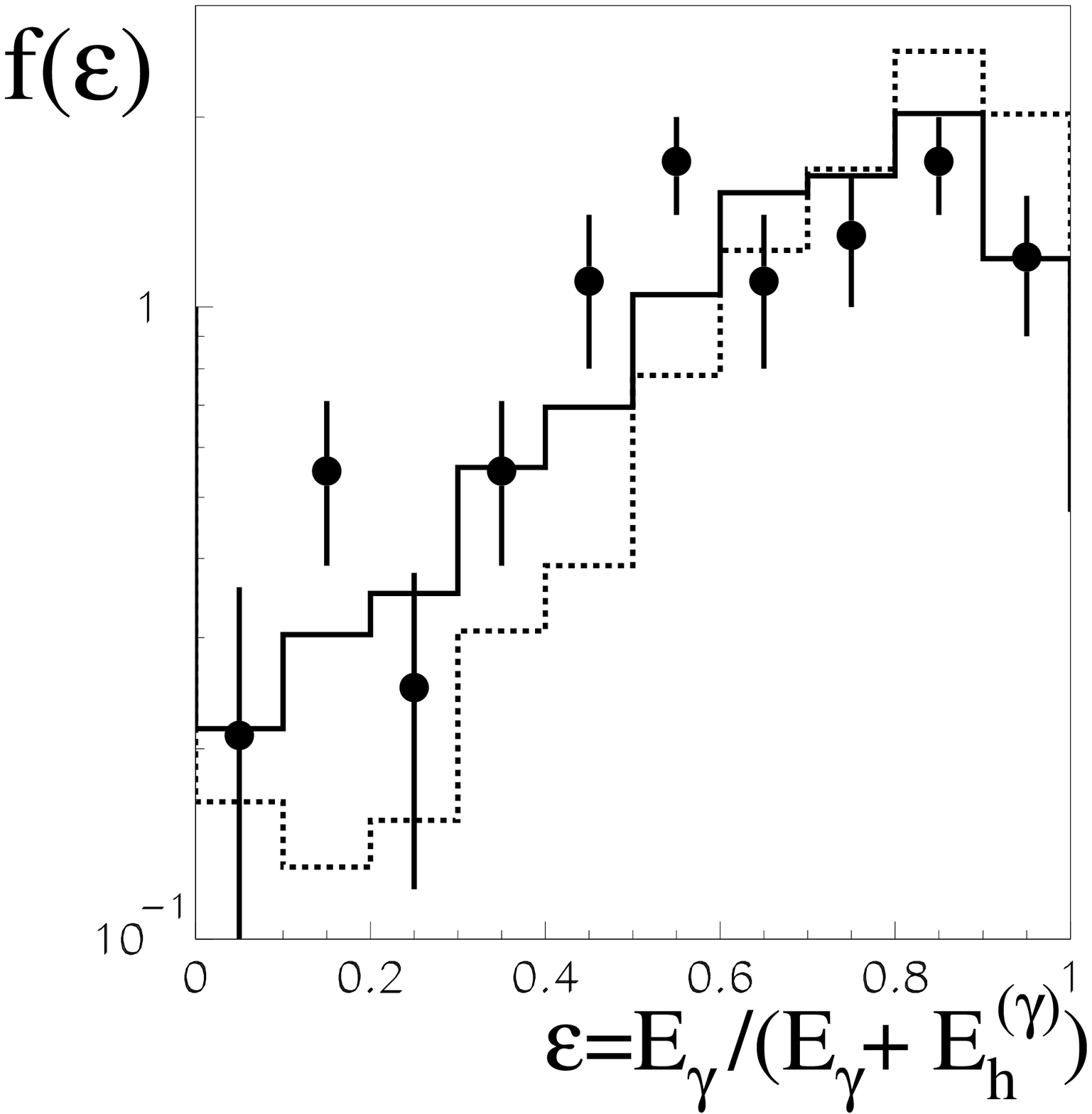}}
\caption{Distribution of electromagnetic energy fraction seen in hadron families at mountain altitude
cosmic ray Chacaltaya experiment \cite{chacaltaya}. The dashed histogram represents FRITIOF model prediction while the solid one 
is the modified version of HDPM algorithm.
\label{eps}}
\end{figure}

It is seen that the data, at least at high $\langle Er\rangle$, can be explained by the simulations. 
The FRITIOF generator does not 
look very well here, but it do not have to be only the problem of transverse momenta. For the
emulsion chamber data the very forward region of particle creation is essential and the 
FRITIOF  was rather tuned (and the ARIADNE \cite{Lonnblad:1999cx} which is the part of 
it responsible for hard gluon bremsstrahlung)
for other sorts of data. This is seen in Fig.\ref{eps} where we plot the distribution of the energy fraction
carried by the energetic photons ($\gamma$~quanta) in the $\gamma$-hadron families observed
by the Chacaltaya experiment.

\section{Conclusions}
We have applied modified, non-extensive statistical thermodynamics to the data on particles with 
high transverse momenta created in high energy hadronic collisions.
It has been found that the data can be successfully described with a fixed temperature parameter $T$ and
a non-extensivity parameter $q$ rising slowly with interaction energy.

The non-extensive thermodynamics can be used to describe phenomenologically the 
long-range correlations (e.g., jet phenomena) without introducing particular physics. 
The achieved agreement for one $q$ value depending only on the interaction energy 
confirms such interpretation of the non-extensivity.

The rise of $q$ leads to systematic rise of the average transverse momentum. The regularity
\begin{displaymath}
q~=~1.25~-~0.33\:s^{-0.054}
\end{displaymath}
allows us to predict with some confidence the rise of $\langle\pt \rangle$ and extrapolate it to the
energies exceeding these available at present accelerators up to the highest cosmic ray energies.
The importance of the exactness of such an extrapolation is clear 
in view of the fact that extensive 
data from the Auger Observatory
are expected soon. The interpretation and understanding 
of EAS features at primary particle energies
exceeding 10$^{20}$eV is a clue to the solution of one of the most exciting astrophysical 
problems.

\bibliography{pt}

\end{document}